\documentclass[iop]{emulateapj}
\usepackage{gensymb}
\usepackage{wasysym}
\usepackage{float}
\usepackage{graphicx}
\usepackage{color}
\usepackage{url}
\usepackage[bottom]{footmisc}
\graphicspath{{images/}}

\newcommand{\jwst}{\textit{JWST}}
\newcommand{\etal}{et~al.}
\newcommand{\Msol}{M$_{\odot}$}

\newcommand{\ebv}{$E(B - V)$}
\newcommand{\ArII}{\hbox{{\rm Ar}\kern 0.1em{\sc ii}}}
\newcommand{\ArIII}{\hbox{{\rm Ar}\kern 0.1em{\sc iii}}}
\newcommand{\CIV}{\hbox{{\rm C}\kern 0.1em{\sc iv}}}
\newcommand{\HI}{\hbox{{\rm H}\kern 0.1em{\sc i}}}
\newcommand{\HII}{\hbox{{\rm H}\kern 0.1em{\sc ii}}}
\newcommand{\HeI}{\hbox{{\rm He}\kern 0.1em{\sc i}}}
\newcommand{\HeII}{\hbox{{\rm He}\kern 0.1em{\sc ii}}}
\newcommand{\NII}{\hbox{{\rm N}\kern 0.1em{\sc ii}}}
\newcommand{\OI}{\hbox{{\rm O}\kern 0.1em{\sc i}}}
\newcommand{\OII}{\hbox{{\rm O}\kern 0.1em{\sc ii}}}
\newcommand{\OIII}{\hbox{{\rm O}\kern 0.1em{\sc iii}}}
\newcommand{\OIIlong}{{\rm O}\kern 0.1em{\sc ii}~$\lambda 3727$} 
\newcommand{\FeII}{\hbox{{\rm Fe}\kern 0.1em{\sc ii}}}
\newcommand{\NeII}{\hbox{{\rm Ne}\kern 0.1em{\sc ii}}}
\newcommand{\NeIII}{\hbox{{\rm Ne}\kern 0.1em{\sc iii}}}
\newcommand{\NeV}{\hbox{{\rm Ne}\kern 0.1em{\sc v}}}
\newcommand{\SII}{\hbox{{\rm S}\kern 0.1em{\sc ii}}}
\newcommand{\SIII}{\hbox{{\rm S}\kern 0.1em{\sc iii}}}
\newcommand{\SIV}{\hbox{{\rm S}\kern 0.1em{\sc iv}}}
\newcommand{\SiIV}{\hbox{{\rm Si}\kern 0.1em{\sc iv}}}
\newcommand{\MgII}{\hbox{{\rm Mg}\kern 0.1em{\sc ii}}}
\newcommand{\Halpha}{\hbox{{\rm H}\kern 0.1em$\alpha$}}
\newcommand{\Hbeta}{\hbox{{\rm H}\kern 0.1em$\beta$}}
\newcommand{\Heopta}{\hbox{{\rm He}\kern 0.1em{\sc i}}~$6678$}
\newcommand{\Heoptb}{\hbox{{\rm He}\kern 0.1em{\sc i}}~$5876$}
\newcommand{\Heoptc}{\hbox{{\rm He}\kern 0.1em{\sc i}}~$4471$}
\newcommand{\Brgam}{\hbox{{\rm Br}\kern 0.1em$\gamma$}}
\newcommand{\Brten}{\hbox{{\rm Br}\kern 0.1em$10$}}
\newcommand{\Breleven}{\hbox{{\rm Br}\kern 0.1em$11$}}
\newcommand{\HeIh}{\hbox{{\rm He}\kern 0.1em{\sc i}}~$1.7$~{\micron}}
\newcommand{\HeIk}{\hbox{{\rm He}\kern 0.1em{\sc i}}~$2.06$~{\micron}}
\newcommand{\squishlist}{
   \begin{list}{$\bullet$}
    { \setlength{\itemsep}{0pt}      \setlength{\parsep}{1pt}
      \setlength{\topsep}{3pt}       \setlength{\partopsep}{0pt}
      \setlength{\leftmargin}{1.5em} \setlength{\labelwidth}{1em}
      \setlength{\labelsep}{0.5em} } }
\newcommand{\squishend}{
    \end{list}  }  
\newcommand{\seventeenlong}  {SDSS~J1723$+$3411}
\newcommand{\twentythreelong}{SDSS~J2340$+$2947}
\newcommand{\seventeen}  {SGAS1723}
\newcommand{\twentythree}{SGAS2340}
\newcommand{\HST}{\textit{HST}}
\newcommand{\JWST}{\textit{JWST}}
\newcommand{\lenstool}{{\tt{Lenstool}}}

\slugcomment{Submitted to ApJ: draft date June 19, 2020}
\shorttitle{Variation in physical conditions within galaxies at cosmic noon}
\shortauthors{Florian et al.}

\begin{document}

\title{Spatial Variation in Strong Line Ratios and Physical Conditions in Two Strongly-Lensed Galaxies at z$\sim$1.4}


\author{Michael K.\ Florian\altaffilmark{1},
Jane R.\ Rigby\altaffilmark{1},
Ayan Acharyya\altaffilmark{2,3},
Keren Sharon\altaffilmark{4},
Michael D.\ Gladders\altaffilmark{5},
Lisa Kewley\altaffilmark{2,3},
Gourav Khullar\altaffilmark{5},
Katya Gozman\altaffilmark{5},
Gabriel Brammer\altaffilmark{6},
Ivelina Momcheva\altaffilmark{7},
David Nicholls\altaffilmark{2,3},
Stephanie LaMassa\altaffilmark{7},  
H\r{a}kon Dahle\altaffilmark{8},
Matthew B.\ Bayliss\altaffilmark{9},
Eva Wuyts\altaffilmark{10},
Traci Johnson\altaffilmark{4}, and
Katherine Whitaker\altaffilmark{11}}

\altaffiltext{1}{Observational Cosmology Lab, NASA Goddard Space Flight Center,
8800 Greenbelt Rd. Greenbelt, MD 20771, USA}

\altaffiltext{2}{Research School of Astronomy and Astrophysics, Australian National University,
Canberra, ACT 2611, Australia}
\altaffiltext{3}{ARC Centre of Excellence for All Sky Astrophysics in 3 Dimensions (ASTRO 3D)}

\altaffiltext{4}{Department of Astronomy, The University of Michigan,
1085 South University Ave, Ann Arbor, MI 48109, USA}

\altaffiltext{5}{Department of Astronomy \& Astrophysics, The University of Chicago, 5640 S.~Ellis Avenue, Chicago, IL 60637, USA}

\altaffiltext{6}{Cosmic Dawn Center, Niels Bohr Institute, University of Copenhagen, Juliane Maries Vej 30, DK-2100 Copenhagen, Denmark}

\altaffiltext{7}{Space Telescope Science Institute, 3700 San Martin Dr., Baltimore, MD 21218, USA}

\altaffiltext{8}{Institute of Theoretical Astrophysics, University of Oslo, P. O. Box 1029, Blindern, N-0315 Oslo, Norway}

\altaffiltext{9}{Department of Physics, University of Cincinnati, Cincinnati, OH 45221, USA}

\altaffiltext{10}{ArmenTeKort, Antwerp, Belgium}

\altaffiltext{11}{Department of Astronomy, University of Massachusetts, Amherst, MA 01003, USA}

\begin{abstract}
For studies of galaxy formation and evolution, one of the major benefits of the James Webb Space Telescope is that space-based IFUs like those on its NIRSpec and MIRI instruments will enable spatially resolved spectroscopy of distant galaxies, including spectroscopy at the scale of individual star-forming regions in galaxies that have been gravitationally lensed.  In the meantime, there is only a very small subset of lensed sources where work like this is possible even with the Hubble Space Telescope's Wide Field Camera 3 infrared channel grisms.  We examine two of these sources, \seventeenlong\ and \twentythreelong, using HST WFC3/IR grism data and supporting spatially-unresolved spectroscopy from several ground-based instruments to explore the size of spatial variations in observed strong emission line ratios like O32, R23, which are sensitive to ionization parameter and metallicity, and the Balmer decrement as an indicator of reddening.  We find significant spatial variation in the reddening and the reddening-corrected  O32 and R23 values which correspond to spreads of a few tenths of a dex in ionization parameter and metallicity.   We also find clear evidence of a negative radial gradient in star formation in \twentythreelong\ and tentative evidence of one in \seventeenlong\, though its star formation is quite asymmetric.  Finally, we find that reddening can vary enough spatially to make spatially-resolved reddening corrections necessary in order to characterize gradients in line ratios and the physical conditions inferred from them, necessitating the use of space-based IFUs for future work on larger, more statistically robust samples.
\end{abstract}


\section{Introduction} \label{sec:intro}
Strong emission line ratios have emerged as powerful diagnostics to understand the physical conditions within distant (redshift $z \ga 1$) galaxies, particularly the metallicity, pressure, and ionization state of the gas, and the amount of dust (e.g., \citealp{Kewley:2002ep}, \citealp{Pettini:2004bq}, \citealp{Rigby:2011il}, \citealp{Steidel:2014es}, \citealp{Nakajima:2014ca}, among many others).  Most observational studies that have used these diagnostics to characterize distant galaxies have, by necessity, used the integrated light of entire galaxies to measure them.  However, a spatially--integrated measurement is unlikely to fairly represent all regions within a galaxy.  Distant galaxies show kiloparsec-scale structures, called ``giant clumps'' (e.g., \citealt{Elmegreen:2005hs, Guo:2012jn}), though recent observational and theoretical results suggest that far smaller spatial scales---tens of parsecs rather than kiloparsecs---are important for star formation in the distant universe \citep{Johnson:2017cd, Mandelker:2013be}.  A spatially-integrated spectrum may well be dominated by the spectrum of one bright giant clump (or a single complex of smaller clumps), if it has extreme line ratios compared to the rest of the galaxy, particularly at bluer rest wavelengths.  For instance, \citet{girard2018} find that about 40\% of the star formation, as indicated by H$\alpha$, lies in just three clumps in a lensed galaxy at $z = 1.6$.  Moving beyond a bulk measurement of galaxy spectra to truly understand the internal physics of these sources requires spectroscopic studies with high spatial resolution.

Surveys using integral field units have characterized how strong line ratios vary spatially in nearby galaxies.  Metallicity gradients, for instance, have been observed in the local universe (e.g., \citealp{belfiore2017}, \citealp{Poetrodjojo:2018fb}), at low redshifts (e.g., $0.1 \leq z \leq\ 0.8$ in \citealp{carton2018}), and at moderate redshifts (e.g., lensed sources at $z = 1.49$ and $2$ in \citealp{Yuan:2011hj} and \citealp{Jones:2010hp}, respectively).  Meanwhile, \citet{Poetrodjojo:2018fb} have investigated, but did not find strong evidence of, radial gradients in ionization parameter at low redshifts, though other studies such as \citet{ellison2018} have found evidence of radial gradients in star formation rate surface densities and \citet{dopita2014} find correlations between SFR and ionization parameter.

While spatial variation of strong line diagnostics at subgalactic scales is well-established in low-redshift galaxies, it is not yet clear how these diagnostics vary spatially in more distant galaxies, which have systematically more extreme physical conditions (\citealp{Kewley:2015kr}, \citealp{holden2016}, \citealp{Onodera:2016aa}) and far more disturbed morphologies in which giant clumpy structures are far more prevalent (\citealp{Elmegreen:2005hs}, \citealp{Shibuya:2016jk}).  Furthermore, the star-formation histories of clumps in such young and disordered objects are likely much different than those of star-forming regions in the local universe.  For example, the star-formation history of a single star-forming clump at $z\sim 1-3$ is more likely to be accurately described by a single stellar population than  structures in older and more morphologically mature galaxies like the Milky Way would be since there has not been as much time for secondary bursts or mixing with older stellar populations from within the same galaxy or from mergers. It is therefore reasonable to suspect that in addition to more extreme physical conditions, higher redshift objects might also have more extreme variations in those conditions.

Additionally, the spatial variation of these diagnostics likely has important consequences for galaxy evolution and cosmology.  For example, the strong line ratio O32 has been found to correlate with Lyman continuum (LyC) leakage \citep{Nakajima:2013ir, Izotov:2018aa}, likely due to its sensitivity to ionization parameter.  But it is unclear whether this is a galaxy-wide phenomenon or a smaller more localized one.  \citet{Izotov:2018aa} and \citet{Nakajima:2013ir} identify LyC leakage in the integrated light of their target galaxies, for example, but in a strongly lensed source, \citet{RiveraThorsen:2019vx} finds LyC leakage from only a single clump.  Understanding the details of processes like this will be critical to understanding how the universe became reionized.

Even with large space telescopes like the Hubble Space Telescope (\HST ) and the upcoming James Webb Space Telescope (\JWST), diffraction limits prevent the study of spatial scales below $\sim 500$~pc in distant field galaxies.  The exceptions are galaxies that have been highly magnified by gravitational lensing, thereby providing access to otherwise inaccessible spatial resolution. Lensing has enabled the measurement of star formation rates \citep{Whitaker:2014jy}, metallicity gradients \citep{Jones:2010hp, Patricio:2019ex}, and rotation curves \citep{Tiley:2019aa, Wuyts:2014eu} with tens of parsecs to a few hundred parsecs resolution for small numbers of galaxies.  

In this Paper, we use \HST\ grism spectroscopy from GO14230 (PI: Rigby) to measure the standard suite of rest-frame optical strong emission lines, from [O~II] 3727, 3729\AA\ to [S~II] 6716, 6733\AA , in two strongly lensed galaxies at z$\sim$1.4 selected from the Sloan Giant Arcs Survey (SGAS).  We map the spatial variation of the diagnostic line ratios H$\alpha$/H$\beta$, R23, O32, and Ne3O2, which are sensitive to dust, metallicity, and ionization parameter.  We quantify the spatial variation in these strong line ratios, and examine the corresponding variation in inferred physical characteristics of the nebular gas.

We also examine whether the spatially integrated spectra of these two galaxies tell the whole story of---or even accurately summarize---the physical properties of the nebular gas in the multiple distinct physical regions that are probed at lensing--boosted spatial resolution.  Each of these objects provides an opportunity to see, directly, how much we miss by using integrated spectra.  \twentythreelong\, at $z=1.42$, is lensed in such a way that there are 4 complete images of the source galaxy, each of which contains three distinct spatial regions from which we can obtain spectra.  The other source, \seventeenlong\ at $z=1.33$, exists in a lensing configuration such that there are two partial (but nearly complete) images of the source combining to form a giant arc as well as two other magnified complete images and a central, demagnified complete image.  The northern most complete image is well-enough separated from the BCG and intracluster light that its spectrum can also be extracted robustly from the \HST\ grism spectroscopy for comparison.

This paper is organized as follows.  \S~\ref{sec:sample} describes the experimental design and sample selection. \S~\ref{sec:data} describes the broad-band imaging, \HST\ spectroscopy, and ground-based spectroscopy that we have obtained to carry out this experiment.  Details of emission line fitting are given in \S~\ref{sec:methods}.  \S~\ref{sec:results} discusses the observed spatial variation of the strong emission line ratios, explores the implied corresponding variation in physical parameters, and compares the implied measurements for spatially integrated versus spatially resolved spectra.  \S~\ref{sec:discussion} discusses the implications of such spatial variation in emission line ratios for studying galaxy evolution and the epoch of reionization, and discusses considerations for future observational and theoretical work.

\section{Experimental Design and Sample Selection}\label{sec:sample}
This study requires technically demanding spectroscopy that fulfills three criteria.  First, the highest possible spatial resolution is needed; this demands that the targets be gravitationally lensed, and further that they be observed either from space or with ground-based adaptive optics systems.  Second, the spectra must have complete wavelength coverage from rest-frame 3727~\AA\ (to cover the [O~II] doublet) to 6730~\AA\ (to cover the [S II] doublet).  Third, the spectra must have excellent relative fluxing over that entire wavelength range, in order to use diagnostic line ratios.  Together, these criteria drive the experiment to use the WFC3-IR grisms onboard \HST .  Of currently available spectrographs, only the WFC3-IR  grisms provide high spatial resolution, excellent flux calibration, and uninterrupted wavelength coverage over this range.  

\begin{figure*}[]
\centering
\includegraphics[width=7in]{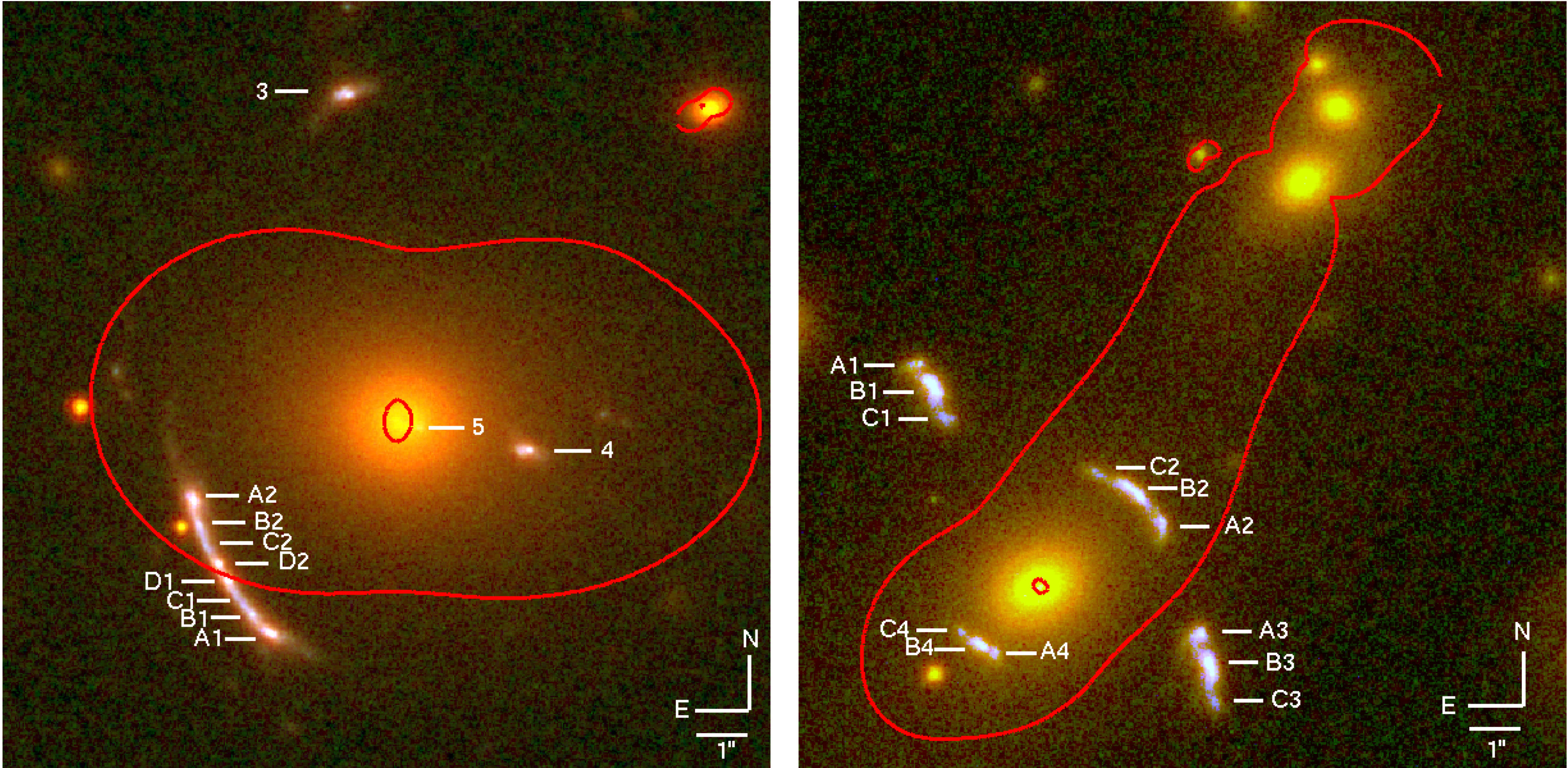}
\caption{Left: Three-color HST image (R: F160W, G:F775W, B:F390W) of \seventeenlong\ with critical curves plotted in red and labels of each image and region in white.  Images 1 and 2 are incomplete images, with regions A--D appearing once in each half of the arc.  Images 3, 4, and 5 are complete images.  Image 3 is separated enough from the BCG that grism spectra could be extracted with minimal contamination.  Right: Three-color HST image (R:F140W, G:F814W, B:F390W) of \twentythreelong\ with critical curves plotted in red and labels for each region in white.  There are four complete images with three distinct physical regions that could be extracted from the grism spectroscopy in each. Lens models are described in section~\ref{sec:lensmodels}.}
\label{fig:critCurves}
\end{figure*}

Used together, the G102 and G141 WFC3-IR grisms can cover the desired range of rest wavelength for galaxies in a narrow redshift range  1.15 $\lesssim z\lesssim$ 1.58.   We therefore selected galaxies in this redshift range from SGAS.  We required that the galaxies appear bright enough to obtain high-quality grism spectra, are highly magnified, and have lensing configurations that are amenable to modeling.  Only two targets emerged from this selection:  \seventeenlong\ (hereafter \seventeen ) with $z = 1.3294 \pm 0.0002$ \citep{Kubo:2010kg}, and \twentythreelong\ (hereafter \twentythree) with $z = 1.420 \pm 0.003$. \seventeen\ is one of the brightest lensed sources in the Sloan Digital Sky Survey (SDSS; \citealp{york:2000sdss}) footprint, and has been reported in several independent searches for lensed galaxies within that data (\citealp{Kubo:2010kg}; \citealp{wen2011}; \citealp{Stark:2013fe}). \twentythree\ has not been previously reported. This lensing system was found as part of the SGAS visual examination of SDSS lines of sight with putative clusters or groups of galaxies, and confirmed as a lens using $gri$ imaging from the 2.5m Nordic Optical Telescope's MOSCA instrument on UT 2012-09-16, with the lensed source then spectroscopically confirmed using the same telescope's ALFOSC spectrograph on UT 2013-09-01.

To support the \HST\ grism spectroscopy, we obtained ground-based spectroscopy from large ground-based telescopes.  We obtained rest-frame ultraviolet (observed-frame optical) spectra with the Echellette Spectrograph and Imager (ESI)  \citep{Sheinis:2002ft} on the Keck~II telescope.  We obtained rest-frame optical spectra of [N~II] and H$\alpha$ with the Gemini Near-InfraRed Spectrograph (GNIRS) \citep{Elias:2006aa} on the Gemini-North telescope.   These instruments provide much higher spectral resolution than the WFC3-IR grisms, but much lower spatial resolution since they are seeing-limited.  Additionally, broadband imaging from \HST\ and Spitzer were leveraged for the creation of lens models, interpretation of source morphologies, and the estimation of stellar masses of each source.

\section{Observations, Data, and Data Reduction}\label{sec:data}
Here we describe the observations, data, and data reduction for the spectra from Keck ESI, Gemini GNIRS, and the \HST\ grisms as well as the broadband imaging from \HST\ and Spitzer.  All wavelengths are listed in vacuum.  Source coordinates and redshifts are listed in Table 1 for ease of reference.

\subsection{Broadband \HST\ Imaging}\label{data-imaging}
We acquired broadband imaging for \twentythree\ and \seventeen\ using the Wide Field Camera 3 (WFC3) onboard \HST , as needed to construct a lens model and do contamination modeling and wavelength calibrations for the grism data.

\seventeen\ was observed in six bands with \HST\ WFC3: F160W, F140W, F110W, and F105W in the IR channel, and F775W and F390W in the UVIS channel.  Imaging in the F160W and F110W bands was conducted as part of \HST\ GO13003 (PI: Gladders) on 2013 March 14 for 1112~s each. Imaging with the UVIS channel in the F775W and F390W bands was conducted as part of the same program, with total integration times of 2380~s and 2368~s respectively.  Imaging in the F140W and F105W bands was conducted as part of \HST\ GO14230 (PI: Rigby) in January and July 2016 alongside the grism observations, for a total integration of 923~s per band. 

\twentythree\ was observed in 4 bands with \HST\ WFC3:  F140W and F105W in the IR channel, paired with the grism observations, and F814W and F390W in the UVIS channel, all from program GO14230 (PI: Rigby).  In the IR channel, \twentythree\ was observed with the F105W filter for 973~s and F140W for 1635~s; in the UVIS channel, it was observed with the F814W filter for 2504~s and F390W for 2600~s.

\begin{table}[h!]
    \label{tab:table1}
    \begin{tabular}{ccccc} 
     \toprule
      Source & RA (J2000) & Dec (J2000) & z$_{gal}$ & $\delta$ z$_{gal}$\\
      \hline
      \seventeen\ & 17:23:36 & +34:11:58 & 1.3293 & 0.0002\\
      \twentythree\ & 23:40:29 & +29:47:47 & 1.42151 & 0.00002 \\
      \hline
    \end{tabular}
        \caption{The positions and redshifts of the targets in this paper.  Right ascensions and declinations (columns 2 and 3) are given in the J2000 coordinates and correspond to the centers of each lensing system (i.e., the BCG).  Redshifts and uncertainties are for the lensed source galaxies and were determined by the GNIRS spectroscopy reported in this paper (\S~\ref{sec:gnirs-results}).}
\end{table}

$\linebreak$
For all \HST\ broad-band imaging, a 4-point dither pattern was used.  To reduce the imaging data, the images were aligned using the Drizzlepac\footnote{http://www.stsci.edu/scientific-community/software/drizzlepac.html} routine \textit{tweakreg}, and drizzled to a common refrence grid with a pixel scale of 0.03\arcsec\ per pixel using \textit{astrodrizzle} with a ``drop size" (final\_pixfrac) of 0.8.  Three-color images of both \seventeen\ and \twentythree\ are shown in Fig.~\ref{fig:critCurves}.

\subsection{Broadband Spitzer Imaging}\label{data-spitzer}
Data from the IRAC instrument of the Spitzer Space Telescope, acquired during the post-cryogenic ``warm mission'', were obtained through program 90232 (for SGAS 1723; PI J.~Rigby) and program 12001 (for SGAS 2340; PI J.~Rigby).  The individual frame times were 30~s; the total per-pixel integration times were 30~min in IRAC channel 1 (3.6 micron), and 10~min in IRAC channel 2 (4.5 micron). 
We processed the Spitzer IRAC Ch1 and Ch2 images as follows.  At a
high level, we followed the general guidance of the IRAC Cookbook for
reducing the COSMOS medium-deep data, though with more stringent (3$\sigma$) outlier rejection, and with residual bias correction.

In more detail, we downloaded the ``corrected basic calibrated data
products'' (cBCDs) from the Spitzer archive.  These cBCDs are the
exposure-level data that have been processed by the IRAC pipeline to
remove instrumental signatures and artifacts, and to calibrate into
physical units.  We applied the warm mission column pulldown
correction (bandcor\_warm by Matt Ashby) to mitigate column artifacts
from bright sources.

For deep integrations, residual bias pattern noise and persistence can
dominate over the background.  To mitigate these effects, we
constructed images of the residual bias, also known as a ``delta dark
frame''.  For each channel in each observation, we created a residual
bias correction from all the cBCDs, by detecting and masking sources
in each image, adjusting the DC level of each image so that the modes
had the same value, and then taking the median with 3$\sigma$
outlier rejection.  The relevant median image was then subtracted from
every cBCD image in that channel and that observation. 

For each target and each filter, we combined the individual images
into a mosaic as follows using the mopex command-line tools.  We used
the overlap correct tool to add an additive correction for each
residual-bias--corrected cBCD image to bring it to a common sky
background level.  We then combined these images into a mosaic using
the mopex mosaic tool, using the drizzle algorithm with a pixel
fraction of 0.6, and 3$\sigma$ outlier rejection using the box outlier
rejection method.

\subsection{Keck ESI spectroscopy}\label{data-esi}
\begin{figure}[]
\includegraphics[width=3.4in]{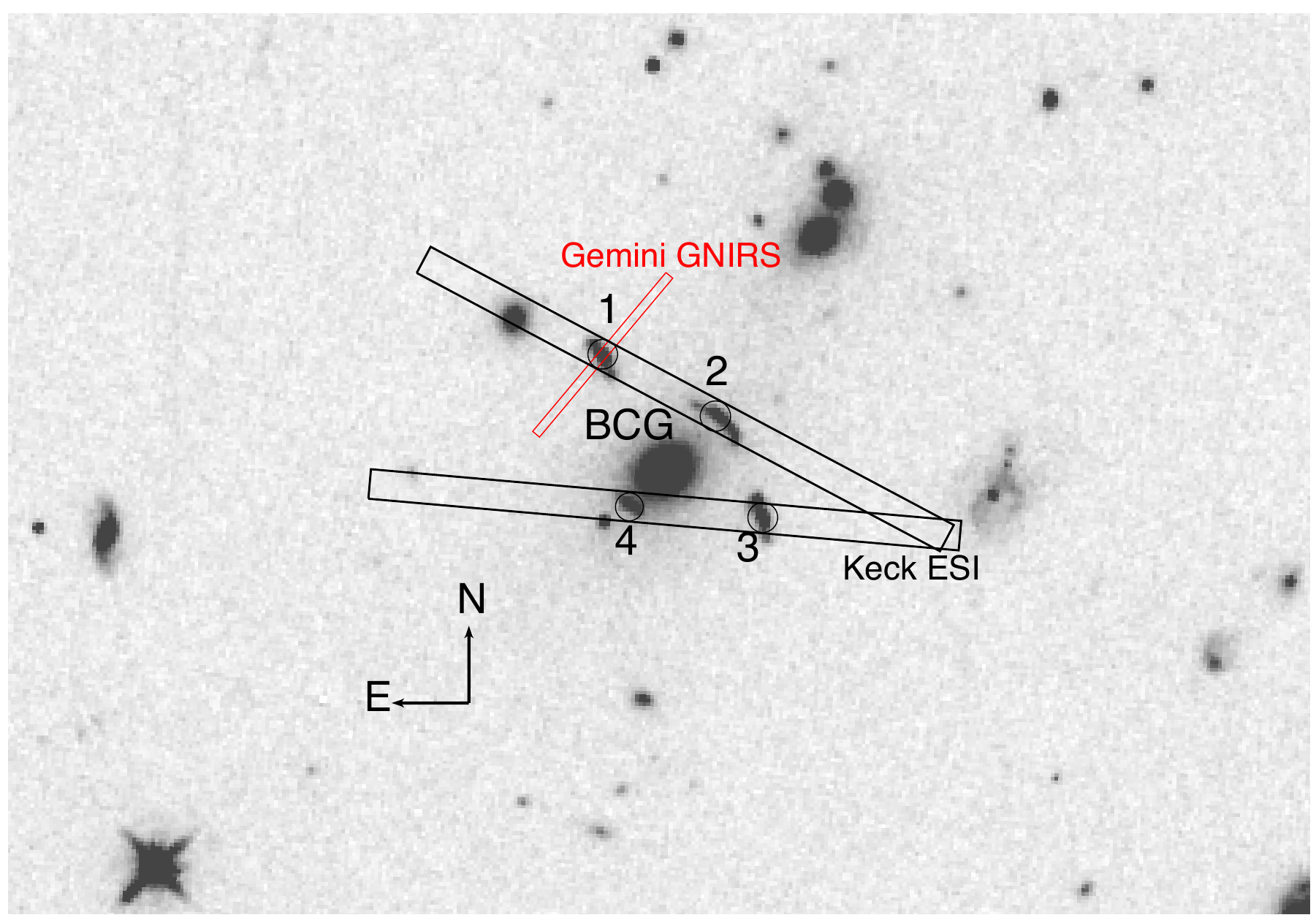}
\caption{Finderchart to show spectroscopic slit positions for the lensed galaxy \twentythreelong .  The background image is \HST\ WFC3-IR F105W.  The long rectangles show the two Keck ESI pointings (with a 1 by 20\arcsec\ long slit), which each captured a pair of images of the lensed galaxy:  1 and 2 in one slit, 3 and 4 in the other.  The red rectangle shows the Gemini GNIRS pointing, with a 0.3 by 7\arcsec\ slit.  The ESI slits were positioned by offsetting from the BCG, which is also marked. North and East are shown with a compass.}
\label{fig:finder2340}
\end{figure}

Both galaxies were observed with the ESI spectrograph on the Keck~II telescope on the nights of 2016 August 27 and 28 UT.   Observing time was obtained through the Australian National University.
Figure~\ref{fig:finder2340} shows how slits were oriented for observations of \twentythree . A similar figure showing slit orientations for observations of \seventeen\ appears in Rigby \etal\ (in prep).  \seventeen\ was observed each night, with a slit placed along the length of the arc, for 3600~s the first night and 3800~s the second night.  For \twentythree, each of the four complete images were observed each night with the following strategy. Images 1 and 2 were observed simultaneously, for 3800~s on the first night and 3600~s on the second night.  Images 3 and 4 were observed simultaneously, for 3600~s on each night.  Light from all three spatial subregions of \twentythree\ was captured by the slit.

We reduced the Keck ESI spectra following the same procedure as Rigby et al.\ 2020 (in prep).  For \twentythree , we extracted the spectrum of each image individually.  Line ratios are measured from the sum of the four images' spectra.

\subsubsection{Gemini GNIRS spectroscopy}\label{data-gnirs}
Gemini GNIRS observations of \seventeen\ and \twentythree\ were obtained through the Gemini Fast Turnaround mode in program GN-2016B-FT-11 (PI Rigby).  Observations were conducted on UT 2016-09-07.   Observations used the short camera, cross-dispersing prism (``SXD'' mode), 0.3\arcsec\ slit, and the 111~lines/mm grating.  Data were obtained as A--B nods.  For \seventeen\ the central wavelength was set to 1.529~$\mu$m; for \twentythree\ it was set to 1.588~$\mu$m.  Figure \ref{fig:finder2340} shows where the GNIRS slit was positioned for \twentythree. A similar figure for \seventeen\ is available in Rigby \etal (in prep). For \seventeen\, the GNIRS slit primarily captured light from the brightest region in the southern part of the arc, region A1. For \twentythree\, the GNIRS slit primarily captured light from the central region, ``B''.  For \seventeen , the cumulative integration time (discarding one exposure with data quality issues) was 2970~s.  For \twentythree, 12 integrations of 270~s duration were obtained, for a total integration time of 3240~s.

The GNIRS spectra were reduced in IRAF using the GNIRS pipeline, which does sky subtraction by differencing A--B pairs.  From each A--B pair, spectra of the A and B images were extracted using a 7~pixel wide boxcar.  We combined the individual spectra by calculating the mean spectrum and the error in the mean.   Rigby \etal\ (in prep) discuss special care that had to be taken with the reduction for \seventeen, due to the fact that the grating shifted position during the observations.

\subsection{\HST\ grism spectroscopy}\label{sec:data-grism}
Spectroscopy for \seventeen\ and \twentythree\ was conducted using the \HST\ WFC3/IR G141 and G102 grisms alongside the direct imaging from program GO14230 described in section~\ref{data-imaging}.

At the redshift of these arcs, z$\sim$1.4, these grism observations cover the wavelength range from just blueward of [O~II] 3727, 3729~\AA\ to just redward of H$\alpha$.  Each target was observed twice with each grism, using two different roll angles to facilitate the modeling of contamination from cluster galaxies.  For \seventeen, observations in each grism were performed for 5112~s at one roll angle, and 4812~s at the other.  \twentythree\ was observed with a similar strategy except that the exposure times were 7518~s for both roll angles with the G141 grism, and 5012 and 5112~s for the two roll angles with the G102 grism.

The \HST\ grism spectra were reduced with the software package Grizli\footnote{https://github.com/gbrammer/grizli}.  We followed the standard Grizli reduction pipeline steps, with a notable extra step of {\tt{GALFIT}}  \citep{Peng:2010eh} modeling to account for the contaminating light of the cluster galaxies.  This extra step is described in detail in Rigby \etal\ (in prep), and was performed on the grism data for both targets. 

In this paper, we separately extract the \HST\ grism spectra of physically distinct regions within each of the lensed galaxies, guided by the source-plane morphologies described in \S~\ref{sec:results-morphologies}.  (This is in contrast to the approach of Rigby \etal\ (in prep), which considers only the spatially-integrated 1D spectrum of the giant arc of \seventeen ). These regions are labeled with letters in Fig.~\ref{fig:critCurves}.  The 1D spectrum of each individual subregion was constructed by summing over the relevant rows in the 2D grism spectrum, and then stacking the spectra of the multiple lensed images of that region.  When the lensing geometry and orientation relative to the dispersion axis of the grisms allows for low morphological broadening and little to no spectral contamination from other pieces of the arcs, spectra from both roll angles are also summed.
 
For \seventeen\ we separately extract the \HST\ grism spectra for the three clumps (A, B, D) and a diffuse component of the giant arc (C), each of which are imaged twice (A1, A2; B1, B2; etc.), and additionally extract the spectrum of the northern complete image (image 3).

For \twentythree\ we separately extract the \HST\ grism spectra for the following physically-distinct regions.  For region A, we extract a spectrum from three of the four lensed images at one roll, and one of the images at the other roll.  For region B, we extract a spectrum from three images at one roll, and two images at the other.  For region C, we extract a spectrum from two images at each roll (one image at both rolls, and two other images at one roll each).  These decisions were guided by the quality of the grism contamination models and the orientation of each image relative to the dispersion axis.  The spectra of each region were stacked over the multiple images and rolls.  Table 2 summarizes which spectra were included in each of these stacks.

The grism spectra were corrected for foreground reddening from the Milky Way galaxy using the \ebv\ values measured by \citet{Green:2015cf}.\footnote{Queried using the python interface provided by those authors at http://argonaut.skymaps.info}

\begin{table}[h!]
    \label{tab:table1}
    \begin{tabular}{ccc} 
     \toprule
      Region & Images From Roll 1 & Images From Roll 2\\
      \hline
      A &  A1, A2, A4 & A1\\
      B &  B1, B2, B4 & B1, B4\\
      C & C2 & C1, C2, C3\\
      Complete Image & 1, 2, 4 & 1\\
      \hline
    \end{tabular}
        \caption{Images included in the stacked spectra for each region of \twentythreelong .  Names correspond to the labels in Fig.~\ref{fig:critCurves}}
\end{table}

\section{Data Analysis Methods}\label{sec:methods}

\subsection{Fitting emission lines in the Keck ESI spectra}\label{methods-esi}
We fit emission lines in the ESI spectra using the continuum and emission line fitting routines described in \citet{ayan:2019}. 

\subsection{Fitting emission lines in the Gemini GNIRS spectra}\label{methods-gnirs}
We fit the spectra with Gaussians using MPFIT, as described in \citet{Wuyts:2014eu}.  The uncertainty spectrum was used as weights in the fitting.  The central wavelengths of all lines were set using the rest wavelength from NIST and  the measured H$\alpha$ redshift. The widths of all lines were forced to vary in lockstep.  The flux ratio of the [N~II] doublet was locked at the value from \citet{Storey:2000jd}.  

\subsection{Fitting emission lines in the \HST\ grism spectra}\label{methods-grism}
Line fluxes were measured from the stacked spectra using the custom fitting technique described in Rigby \etal\ (in prep), with a few minor changes to compensate for the lower signal-to-noise ratios of the spectra of individual regions, compared to the spatially--integrated spectrum of the giant arc studied in that paper.  Two iterations of the fitting algorithm were run on each spectrum.  The first iteration solves for the redshift, the morphological line broadening (a nuisance parameter), and the line fluxes and uncertainties.  The second iteration fixes the redshift and morphological broadening parameter determined in the first iteration, and then allows for small variations in the observed wavelength of each emission feature; this is motivated by known uncertainties in the wavelength solutions of \HST\ grism spectra.  The second iteration typically resulted in better overall spectral fits.  However, for a few low signal-to-noise spectra (the integrated spectra of images 1, 2, and 3 of \twentythree\ from the G102 grism), the second iteration resulted in poorer fits; for those spectra we report only the line fluxes measured from the first iteration.

At the low spectral resolution of the WFC3/IR grisms, the [N~II] doublet is unrecoverably blended with H$\alpha$.  Therefore, we set the [N~II] doublet ratio to its theoretical value \citep{Storey:2000jd}, and set the [N~II]/H$\alpha$ ratio to the value measured from the spatially--integrated Gemini GNIRS spectrum of each galaxy (see \S\ref{sec:gnirs-results}).  Similarly, the [O~II]~3727/3729 flux ratio was set to the value measured from the Keck ESI spectra (\S~\ref{sec:esi-results}).  These values were used for all physical regions of \seventeen\ and \twentythree\ despite coming from the integrated spectrum.

\section{Results}\label{sec:results}

\subsection{Lens Models}\label{sec:lensmodels}
We used the strong lensing evidence, i.e., identification of multiply-imaged lensed galaxies, in order to 
compute strong lensing models for \seventeen\ and \twentythree. The lensing analysis of \seventeen\ was published in \citet{masstadon} as part of the SGAS-HST program. We describe here the lensing analysis of \twentythree . However, we note that the lensing analysis of both clusters follows the same procedure.

We identified two strongly lensed galaxies in this field. Source~1, the topic of this paper, has four complete images
around the core of the cluster. The images are resolved and show identical morphology and color variation. We used the central positions of four distinct morphological features in each image, as well as the spectroscopic redshift of the source galaxy, $z_{spec}=1.42$, as lensing constraints. Source~2 appears as three images of a faint galaxy, $6\farcs44$ northwest of the BCG. Its redshift is unknown, and was used as a free parameter in the lens modeling process.  Constraints are tabulated in Table 3

The lens model was computed using the publicly available software \lenstool\ \citep{jullo07}. \lenstool\ uses a bayesian approach to explore the parameter space and identify the best-fit model, as the set of parameters that minimize the scatter between the predicted and observed lensed images in the image plane. The lens model results in a parametric two dimensional description of the lens plane, from which the projected mass density, lensing magnification, and deflection are derived. 

The lens was modeled as a linear combination of parameterized mass halos. Each halo was assumed to be a Pseudo Isothermal Elliptical Mass Distribution \citep[PIEMD, or dPIE;][]{eliasdottir07}, with the following parameters: position $x$, $y$; ellipticity $e$; position angle $\theta$; core radius $r_c$; truncation radius $r_{cut}$, and normalization $\sigma_0$. The lens is composed of cluster-scale halos and galaxy-scale halos. The latter were placed at the positions of observed cluster-member galaxies, identified using the red sequence technique \citep{gladders00} in the F814W-F105W color-magnitude space. The position, ellipticity, and position angle of the galaxy halos were fixed to their observed values as measured with Source Extractor \citep{bertin96}. The other parameters are scaled to the luminosity, following \citet{limousin05}.
All the parameters of cluster-scale halos were left free, except for $r_{cut}$ which is beyond the range that can be constrained by the lensing evidence. It was fixed to 1500 kpc.

We find that the lens is best described by two cluster/group scale halos, combined with galaxy scale halos.  The statistical uncertainties were derived by computing the lensing outputs from sets of parameters drawn from steps in the MCMC chain.  Magnifications and the corresponding uncertainties are tabulated for both this source and \seventeen\ in Table 4. The relatively small number of lensed galaxies in this field limited the number of constraints available for lens modeling, resulting in more uncertain models and relatively larger statistical uncertainties in magnification than for \seventeen. 

\begin{table}[h!]
        \label{tab.arcstable}
        \begin{tabular}{llllll} 
        \toprule
        \colhead{ID} &
        \colhead{R.A.}    & 
        \colhead{Decl.}    & 
        \colhead{$z_{spec}$}     & 
        \colhead{$z_{model}$}   \\[-0.1cm]
        \colhead{} &
        \colhead{(J2000)}     & 
        \colhead{(J2000)}    & 
        \colhead{}       & 
        \colhead{}       \\     
        \hline
A1 & 355.119674 & 29.797717 &  & \\
A2 & 355.118160 & 29.796868 &  & \\
A3 & 355.117915 & 29.796249 &  & \\
A4 & 355.119201 & 29.796155 &  & \\
B1a & 355.119622 & 29.797621 & 1.4200 & \\
B2a & 355.118294 & 29.796994 &  & \\
B3a & 355.117874 & 29.796099 &  & \\
B4a & 355.119274 & 29.796182 &  & \\
B1b & 355.119573 & 29.797569 &  & \\
B2b & 355.118377 & 29.797054 &  & \\
B3b & 355.117848 & 29.796033 &  & \\
B4b & 355.119316 & 29.796206 &  & \\
C1 & 355.119524 & 29.797421 &  & \\
C2 & 355.118555 & 29.797102 &  & \\
C3 & 355.117837 & 29.795886 &  & \\
C4 & 355.119419 & 29.796267 &  & \\
\hline
11.1 & 355.118291 & 29.798323 & & $1.34^{+0.68}_{-0.42}$ & \\
11.2 & 355.117948 & 29.798055 &  & & \\
11.3 & 355.117203 & 29.797563 &  & & \\
\hline
\end{tabular}
 \caption{Lensing constraints used in the lens model of \twentythreelong. IDs correspond to labels in Fig. ~\ref{fig:critCurves}. Region B has been broken up into two components (labeled with lowercase a and b) for modeling.  B1a is the northern half of region B in image 1.  B2a is the southern half of region B in image 1.  IDs beginning with ``11" correspond to a second source not studied in this paper.}
\end{table}
\begin{table*}[h!]
    \centering
    \label{tab:magn}
    \begin{tabular}{cccccccccccc} 
     \toprule
      Source & Region & Magnification & $m_{160}$ & $m_{140}$ & $m_{110}$ & $m_{105}$ & $m_{814}$ & $m_{775}$ & $m_{390}$ & $fH\beta\ _{obs}$ & $\delta\ fH\beta\ _{obs}$\\
     \hline
     & & & & & & & & & & & \\[-0.15cm]
     1723+3411 & A1 & $16.4^{+0.5}_{-1.4}$ & 21.8 & 21.9 & 21.9 & 21.5 & -- & 22.3 & 22.4 & 15.81 & 0.75  \\[0.2cm]
     1723+3411 & A2 & $16.6^{+0.5}_{-1.6}$ & 21.9 & 22.1 & 22.0 & 21.7 & -- & 22.4 & 22.6 & 18.81 & 0.63 \\[0.2cm]
     1723+3411 & B1 & $25.0^{+0.9}_{-2.1}$ & 22.6 & 22.4 & 22.8 & 22.2 & -- & 22.9 & 22.9 & 7.01 & 0.35 \\[0.2cm]
     1723+3411 & B2 & $25.9^{+0.9}_{-2.3}$ & 22.5 & 22.5 & 22.5 & 22.2 & -- & 22.9 & 23.0 & 7.07 & 035 \\[0.2cm]
     1723+3411 & C1 & $36.2^{+0.9}_{-3.3}$ & 23.1 & 23.0 & 23.3 & 22.8 & -- & 23.3 & 23.3 & 4.76 & 0.35 \\[0.2cm]
     1723+3411 & C2 & $40.3^{+1.4}_{-3.6}$ & 23.3 & 23.2 & 23.4 & 23.0 & -- & 23.7 & 23.6 & 6.32 & 0.46 \\[0.2cm]
     1723+3411 & D1 & $102.7^{+8.2}_{-8.5}$ & 22.2 & 22.3 & 22.4 & 22.2 & -- & 22.7 & 23.0 & 5.29 & 0.47 \\[0.2cm]
     1723+3411 & D2 & $112.6^{+7.6}_{-10.1}$ & 22.3 & 22.2 & 22.4 & 22.2 & -- & 22.6 & 22.9 & 4.78 & 0.43 \\[0.2cm]
     1723+3411 & Giant Arc & $52.7^{+3.3}_{-1.2}$ & 20.1 & 20.2 & 20.2 & 19.9 & -- & 20.5 & 20.6 & 71.14 & 2.09 \\[0.2cm]
     1723+3411 & Complete Image 3& $7.2_{-0.31}^{+0.33}$ & 22.0 & 22.1 & 22.0 & 21.8 & -- & 22.5 & 23.6 & 10.72 & 0.90 \\[0.2cm]
     2340+2947 & A1 & $8.9^{+9.0}_{-2.2}$ & -- & 23.3 & -- & 23.8 & 23.8 & -- & 25.1 & 6.50 & 0.45 \\[0.2cm]
     2340+2947 & A2 & $8.5^{+8.3}_{-1.7}$ & -- & 22.9 & -- & 23.4 & 23.4 & -- & 24.4 & 9.49 & 0.65 \\[0.2cm]
     2340+2947 & A3 & $12.8^{+12.2}_{-3.2}$ & -- & 22.7 & -- & 23.2 & 23.2 & -- & 24.1 & 12.47 & 0.86 \\[0.2cm]
     2340+2947 & A4 & $3.1^{+3.0}_{-0.7}$ & -- & 23.5 & -- & 24.0 & 24.1 & -- & 25.2 & 3.94 & 0.27 \\[0.2cm]
     2340+2947 & B1 & $9.8^{+10.0}_{-2.2}$ & -- & 22.1 & -- & 22.4 & 22.5 & -- & 23.3 & 15.51 & 0.95 \\[0.2cm]
     2340+2947 & B2 & $6.5^{+6.4}_{-1.5}$ & -- & 22.2 & -- & 22.5 & 22.6 & -- & 23.5 & 12.72 & 0.78 \\[0.2cm]
     2340+2947 & B3 & $9.2^{+8.1}_{-2.3}$ & -- & 22.1 & -- & 22.4 & 22.5 & -- & 23.3 & 15.04 & 0.92 \\[0.2cm]
     2340+2947 & B4 & $2.8^{+2.5}_{-0.6}$ & -- & 22.8 & -- & 23.2 & 23.3 & -- & 24.3 & 5.82 & 0.36 \\[0.2cm]
     2340+2947 & C1 & $11.4^{+11.3}_{-2.5}$ & -- & 23.6 & -- & 23.8 & 24.0 & -- & 24.7 & 7.85 & 0.68 \\[0.2cm]
     2340+2947 & C2 & $6.8^{+6.9}_{-1.7}$ & -- & 23.5 & -- & 23.9 & 24.1 & -- & 25.1 & 5.57 & 0.48 \\[0.2cm]
     2340+2947 & C3 & $7.6^{+7.0}_{-1.8}$ & -- & 23.8 & -- & 24.2 & 24.3 & -- & 25.0 & 5.81 & 0.50 \\[0.2cm]
     2340+2947 & C4 & $3.2^{+2.7}_{-0.7}$ & -- & 24.4 & -- & 24.8 & 25.0 & -- & 26.1 & 2.16 & 0.19 \\[0.2cm]
     2340+2947 & Complete Image 1 & $9.9^{+9.9}_{-2.3}$ & -- & 21.6 & -- & 22.0 & 22.0 & -- & 22.9 & 30.60 & 1.77 \\[0.2cm]
     2340+2947 & Complete Image 2 & $7.1^{+7.0}_{-1.6}$ & -- & 21.6 & -- & 21.9 & 22.0 & -- & 22.9 & 27.72 & 1.61 \\[0.2cm]
     2340+2947 & Complete Image 3 & $9.7^{+8.5}_{-2.4}$ & -- & 21.5 & -- & 21.8 & 21.9 & -- & 22.7 & 33.24 & 1.93 \\[0.2cm]
     2340+2947 & Complete Image 4 & $3.0^{+2.7}_{-0.7}$ & -- & 22.2 & -- & 22.6 & 22.7 & -- & 23.8 & 12.13 & 0.70 \\[0.2cm]
    \end{tabular}
        \caption{Magnifications, magnitudes, and H$\beta$ fluxes of each of the images and subregions of \seventeen\ and \twentythree .  Columns 1 and 2 denote the source and the image or subregion.  Column 3 contains the magnifications of each region and each complete image of \seventeen\ and \twentythree.  Uncertainties are reported as the 68\% level. e.g., for \seventeen\ region A1, 68\% of models that produce magnifications below 16.6 produce magnifications above 15.0 and 68\% of models that produce magnifications larger than 16.6 produce magnifications below 17.1.  Columns 4-10 contain the magnitudes of each region or complete image obtained from simple aperture photometry in each of the available \HST\ filters.  For the purposes of planning future observations, the observed H$\beta$ flux is included for each component in column 11 with its uncertainty in column 12.  Rather than reporting line fluxes based on fits of individual, often low S/N spectra, we report values measured from the stacked spectra of all images of each region, scaled based on the number of roll angles and the flux ratios from the broadband imaging.  The flux ratios used for scaling here are determined from the stacks of the imaging data in the UVIS channel because the higher spatial resolution allows better separation of the subregions while the flux ratios are consistent regardless of wavelength.  A full list of fluxes, scaled to H$\beta$ is available as a machine-readable table.}
\end{table*}

\subsection{Source Morphologies}\label{sec:results-morphologies} 
The lens models enable a reconstruction of the source-plane morphology, which informs our extraction and summation of the \HST\ grism data.  

The giant arc in \seventeen\ is a merging pair of images of a single source galaxy.  Each of these images is nearly complete.  As Figure~\ref{fig:critCurves} shows, six large clumps are apparent in the broadband imaging; these are two images of each of the three physically distinct clumps (A, B, and D).  An additional region, labeled C, appears to correspond to either a diffuse component, or to multiple spatially un-resolved clumps.  In addition to the giant arc in \seventeen , there are two complete but less magnified images of the lensed galaxy (images 3 and 4) and a demagnified central image (5).  The grism spectra of images 4 and 5 are badly contaminated by the BCG; we were, howerver, able to extract the spectrum of the northern complete image (3).

There are four lensed images of the source galaxy \twentythree.  Each features a bright central region (B) and two fainter clumps near the edges (A and C).  All of these regions, for both sources, are labeled in Figure~\ref{fig:critCurves} and correspond to the regions from which the spectra described in \S~\ref{sec:data-grism} were extracted.  Because the dither pattern of the \HST\ imaging observations allows us to oversample the PSF, our final drizzled data products are able to resolve region B into two sub-components for the purposes of lens modeling.  However, drizzling spectra produces correlated noise which can mimic emission or absorption features, so we are forced to use coarser pixels for the spectral data and interlace rather than drizzle, which reduces the effective spatial resolution of the spectral data and blends the spectra of the two sub-components of region B.

To summarize, for \seventeen\ we extract spectra of four physically distinct regions as well as one complete image; for \twentythree\ we extract spectra of three regions and also create a stacked composite spectrum of the complete images.

\subsection{Photometry from the \HST\ images}\label{sec:results-photometry}

Broadband magnitudes for \twentythree\ and the giant arc in \seventeen\ were determined using a {\tt{GALFIT}} decomposition of each source galaxy.  Using these integrated magnitudes, magnitudes measured from the Spitzer data, the magnifications in Table 4 and the stellar population synthesis parameter inference code, {\tt{Prospector}} \citep{leja:2017a}, we estimated stellar masses of $9.67^{+9.0}_{-4.4}  \times 10^{8}$ \Msol\ for the source in \twentythree\ and $5.95^{+2.2}_{-1.86} \times 10^{8}$ \Msol\ for \seventeen\ .  Details of these measurements and the process used to infer the stellar masses are included in the Appendix.  For the purposes of planning future observations, approximate apparent magnitudes (uncorrected for lensing magnification) of each individual region in these images based on photometry from custom apertures are included, along with the magnifications, in Table 4.  We also include the estimated H$\beta$ flux for each region determined by correcting the H$\beta$ flux from the stacked spectrum using the observed broadband flux ratios of the various images of each region.

\subsection{Keck ESI results}\label{sec:esi-results}
 Fluxes for emission lines measured from the ESI spectra for \seventeen\ are listed in Table 1 of Rigby \etal\ (in prep), and for \twentythree\ are listed in Table 5 of this paper.

For \seventeen , the [O~II]~3727/3729 flux ratio was measured as $0.71 \pm 0.01$.
For \twentythree, the [O~II]~3727/3729 flux ratio was measured as $0.82 \pm 0.01$.

These values are measured from spatially integrated spectra. Because the \HST\ grisms do not spectrally resolve this doublet, the pressures assumed in section~\ref{sec:gridstuff} are based on these values.

\begin{figure}[]
\centering
\includegraphics[width=3.4in]{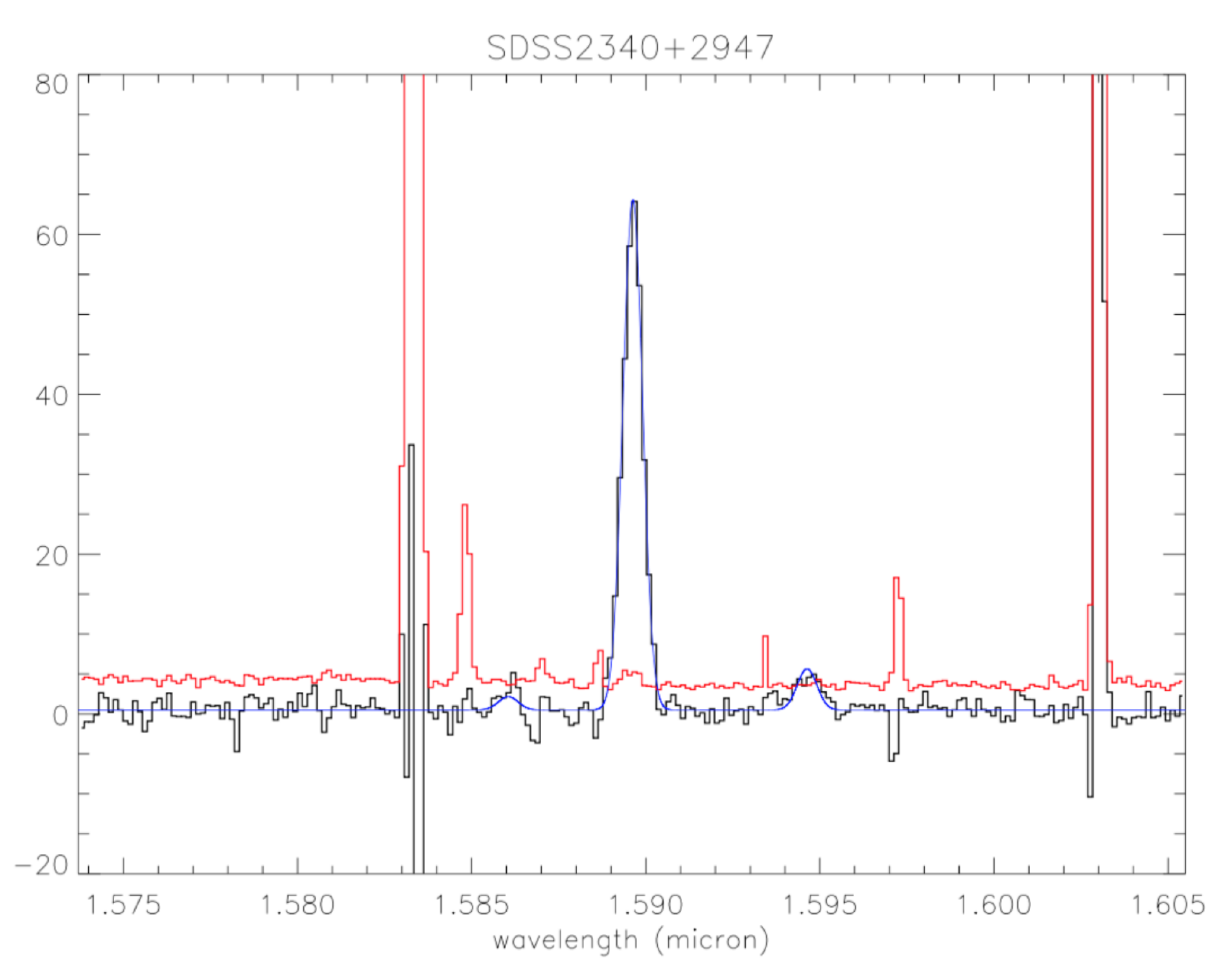}
\caption{The $\alpha$ and the [N~II] doublet in the Gemini GNIRS spectrum of \twentythree .  Black steps show the summed one-dimensional spectrum; red steps show the uncertainty spectrum, and the blue curve shows the best fit.}
\label{fig:ha2340}
\end{figure}

\subsection{GNIRS Gemini results}\label{sec:gnirs-results}
For \seventeen , results from the Gemini GNIRS spectrum were presented in Rigby \etal\ (in prep). 
The redshift was measured as  $z(H\alpha) =  1.3293 \pm 0.0002$.
The [N~II] 6586 / H$\alpha$ flux ratio was measured to be $0.062  \pm  0.011$. 

For \twentythree , we measure the redshift of H$\alpha$ as  $z = 1.42151 \pm 0.00002$ .
The [N~II] 6586 / H$\alpha$ flux ratio is measured to be $0.084 \pm 0.03$ (Fig.~\ref{fig:ha2340}).

These values are spatially integrated for \twentythree\ and come from clump A1 in \seventeen\ .

\subsection{Line flux measurements from the \HST\ grism spectra}\label{sec:results-grismfluxes}

\begin{figure}[]
\includegraphics[width=3.4in]{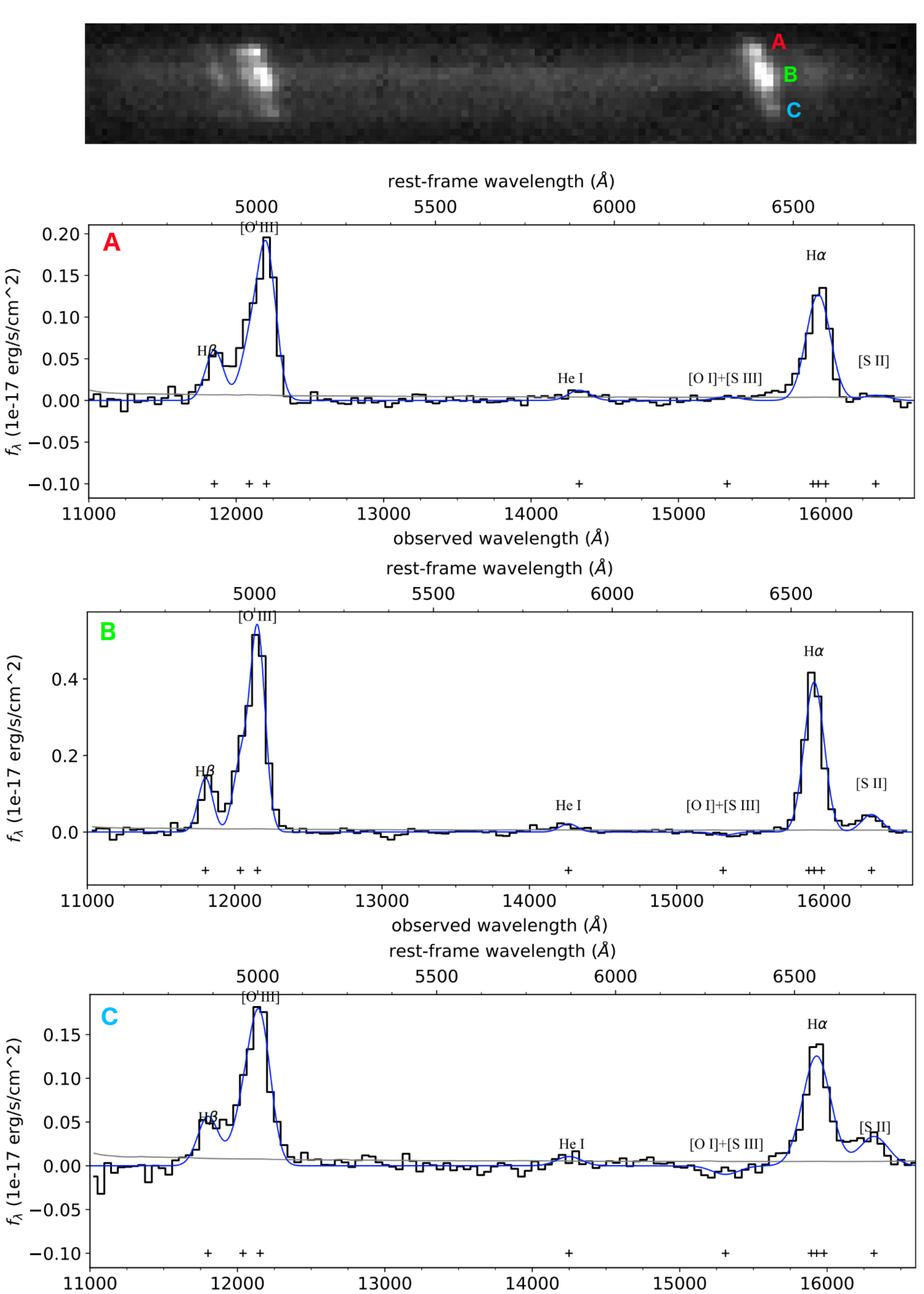}
\caption{An example of a reduced 2D grism spectrum (top) for \twentythree\ and the extracted 1D spectra for each of three regions labeled A (second from top), B (second from bottom), and C (bottom), corresponding to the labels in Fig.~\ref{fig:critCurves}.  The 1D continuum-subtracted spectrum is plotted in black, with the fit in blue.  The continuum fit is plotted in grey.  Emission lines are labeled and their best-fit wavelengths are marked by crosses.  This example uses the G141 grism.}
\label{fig:grisExample}
\end{figure}

An example of a reduced 2D grism spectrum for a single image of \twentythree\ and the 1D extraction and linefitting for each subregion is shown in Fig~\ref{fig:grisExample}.  Line fluxes for the stacks of each region in \seventeen\ and \twentythree\ normalized to the H$\beta$ flux, as well as line fluxes for each individual image of each region normalized to the H$\beta$ flux are included as machine readable tables in the electronic version of this article. These are included with the understanding that they may be needed for planning future observations and therefore the H$\beta$ fluxes are reported as observed, not corrected for the lensing magnification.  Corrections can be made using the information in Table 4.

\subsection{Spatial variation of strong line diagnostics}
We investigate the spatial variation of  several strong line diagnostics sensitive to dust, ionization parameter, metallicity and star formation rate.  Of these, only the H$\alpha$ flux, a star formation rate indicator, is sensitive to the magnification due to gravitational lensing.  The others, because they are flux ratios rather than fluxes, are invariant under lensing.  Here, we describe the degree of variation seen in observables---line fluxes and ratios.  In \S~\ref{sec:results-logUandZ} we explore how these variations correspond to variation in physical parameters, namely metallicity and ionization parameter.

\subsubsection{Ratios Sensitive to Reddening: H$\alpha$/H$\beta$ and H$\beta$/H$\gamma$}
We use Balmer decrements to estimate the reddening at each region within these two sources.
For \seventeen , H$\alpha$ falls in G141, H$\gamma$ falls in G102, and H$\beta$ is covered by both grisms.  We find that H$\beta$ fluxes measured in G102 are more consistent from roll-to-roll than those measured from G141; we attribute this to H$\beta$ falling at an observed wavelength where the sensitivity of the G141 grism is rapidly declining.  For \twentythree, H$\alpha$ and H$\beta$ are both captured by the G141 grism, while H$\gamma$ is captured by G102.

\begin{figure}[]
\includegraphics[width=3.4in]{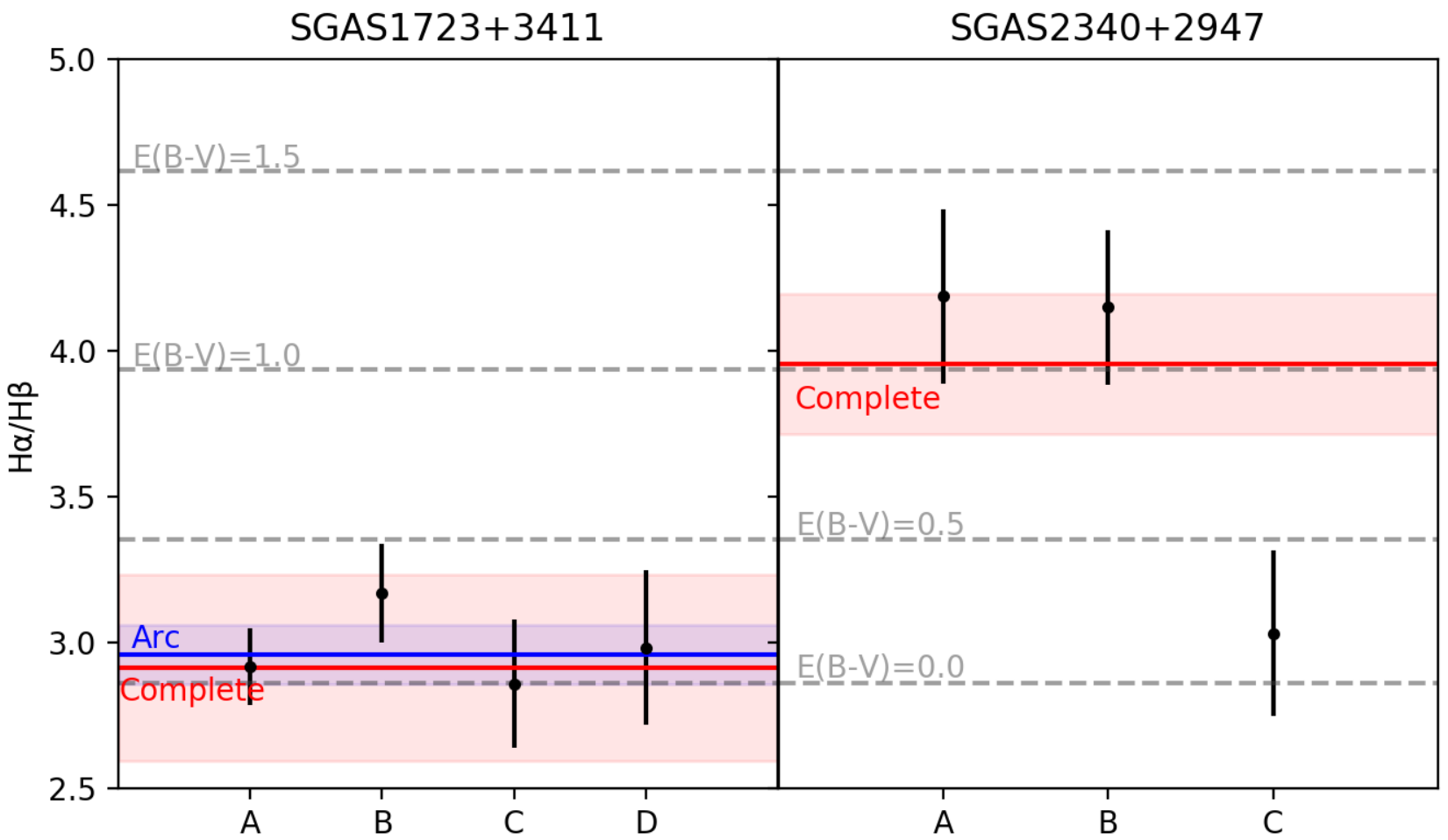}
\caption{The H$\alpha$ to H$\beta$ ratios in \seventeen\ (left) and \twentythree\ (right).  The latter shows significant spatial variation, with region C being uniquely low.  The spatial variation in \seventeen\ is less pronounced, though the difference between region B and region C is suggestive (but not statistically significant). The red bands in each plot show the spatially integrated values measured from the stack of the complete images of \twentythree , and from the northern complete image of \seventeen ; the widths of the bands correspond to the uncertainties.  In blue bands we show the values measured from the integrated spectrum of the giant arc in \seventeen .  For reference, several values of \ebv for case B recombination and R$_{v}$=3.1 are shown by gray dashed lines.   While all of the H$_{\alpha}$/H$_{\beta}$ values are in general agreement for \seventeen , it is clear that the value inferred from the integrated spectrum of \twentythree\ is not representative of the reddening of region C, while the contribution of region C also pulls the reddening of the complete image down relative to regions A and B.  This is discussed in section~\ref{sec:reddening_var_imp}}
\label{fig:reddeningWBands}
\end{figure}

Fig.~\ref{fig:reddeningWBands} shows the measured H$\alpha$/H$\beta$ flux ratios.  From these, we infer E(B-V) reddening values assuming case B recombination.  

\begin{table*}[h]
\centering
\begin{tabular}{lrlllll}
\toprule
         Line ID &  $\lambda_{\mathrm{rest}}$ & W$_{\mathrm{r,fit}}$ & $\delta$ W$_{\mathrm{r,fit}}$ & W$_{\mathrm{r,signi}}$ & flux$_{\mathrm{obs}}$ & $\delta$ flux$_{\mathrm{obs}}$ \\
& (\AA) & (\AA) & (\AA) & (\AA) & (10$^{-17}$ ergs/s/cm$^2$) & (10$^{-17}$ ergs/s/cm$^2$) \\
\hline
     O III] 1660 &                  1660.8090 &               -3.05 &                          1.86 &                   4.17 &                 28.6 &                          17.39 \\
     O III] 1666 &                  1666.1500 &           $>$-2.04 &                            .. &                     .. &              $<$8.38 &                             .. \\
     N III] 1750 &                  1749.7000 &             $>$-0.77 &                            .. &                     .. &              $<$7.97 &                             .. \\
  {[}Si II] 1808 &                  1808.0130 &           $>$-4.29 &                            .. &                     .. &              $<$44.4 &                             .. \\
  {[}Si II] 1816 &                  1816.9280 &          $>$-36.6 &                            .. &                     .. &             $<$376. &                             .. \\
 {[}Si III] 1882 &                  1882.7070 &           $>$-0.37 &                            .. &                     .. &               $<$3.13 &                             .. \\
    Si III] 1892 &                  1892.0290 &           $>$-0.33 &                            .. &                     .. &               $<$2.72 &                             .. \\
  {[}C III] 1906 &                  1906.6800 &           $>$-0.30 &                            .. &                     .. &               $<$2.46 &                             .. \\
     C III] 1908 &                  1908.7300 &           $>$-0.29 &                            .. &                     .. &               $<$2.37 &                             .. \\
      N II] 2140 &                  2139.6800 &           $>$-0.37 &                            .. &                     .. &               $<$0.69 &                             .. \\
  {[}O III] 2320 &                  2321.6640 &           $>$-0.33 &                            .. &                     .. &               $<$0.59 &                             .. \\
      C II] 2323 &                  2324.2140 &           $>$-0.33 &                            .. &                     .. &               $<$0.60 &                             .. \\
     C II] 2325c &                  2326.1130 &              -0.98 &                          0.44 &                   8.81 &                  4.28 &                           1.92 \\
     C II] 2325d &                  2327.6450 &           $>$-0.32 &                            .. &                     .. &               $<$0.58 &                             .. \\
      C II] 2328 &                  2328.8380 &            $>$-0.33 &                            .. &                     .. &               $<$0.59 &                             .. \\
    Si II] 2335a &                  2335.1230 &           $>$-0.32 &                            .. &                     .. &               $<$0.58 &                             .. \\
    Si II] 2335b &                  2335.3210 &           $>$-0.33 &                            .. &                     .. &               $<$0.59 &                             .. \\
      Fe II 2365 &                  2365.5520 &               -0.45 &                          0.22 &                   3.54 &                  1.96 &                           0.98 \\
     Fe II 2396a &                  2396.1497 &           $>$-0.49 &                            .. &                     .. &               $<$0.89 &                             .. \\
     Fe II 2396b &                  2396.3559 &           $>$-0.49 &                            .. &                     .. &               $<$0.89 &                             .. \\
   {[}O II] 2470 &                  2471.0270 &               -0.52 &                          0.22 &                   3.46 &                  2.19 &                           0.92 \\
      Fe II 2599 &                  2599.1465 &           $>$-0.34 &                            .. &                     .. &               $<$0.56 &                             .. \\
      Fe II 2607 &                  2607.8664 &           $>$-0.38 &                            .. &                     .. &               $<$0.63 &                             .. \\
      Fe II 2612 &                  2612.6542 &           $>$-0.40 &                            .. &                     .. &               $<$0.65 &                             .. \\
      Fe II 2614 &                  2614.6051 &           $>$-0.47 &                            .. &                     .. &               $<$0.77 &                             .. \\
      Fe II 2618 &                  2618.3991 &           $>$-0.43 &                            .. &                     .. &               $<$0.69 &                             .. \\
      Fe II 2621 &                  2621.1912 &           $>$-0.41 &                            .. &                     .. &               $<$0.67 &                             .. \\
      Fe II 2622 &                  2622.4518 &           $>$-0.49 &                            .. &                     .. &               $<$0.8 &                             .. \\
      Fe II 2626 &                  2626.4511 &              -0.65 &                          0.21 &                   4.33 &                  2.56 &                           0.84 \\
      Fe II 2629 &                  2629.0777 &           $>$-0.62 &                            .. &                     .. &               $<$1.01 &                             .. \\
      Fe II 2631 &                  2631.8321 &           $>$-0.47 &                            .. &                     .. &               $<$0.76 &                             .. \\
      Fe II 2632 &                  2632.1081 &           $>$-0.46 &                            .. &                     .. &               $<$0.75 &                             .. \\
     Mg II 2797b &                  2798.7550 &              -1.40 &                          0.15 &                  13.8 &                 5.33 &                           0.57 \\
     Mg II 2797d &                  2803.5310 &              -0.85 &                          0.15 &                   7.96 &                  3.25 &                           0.57 \\
       He I 2945 &                  2945.1030 &           $>$-0.47 &                            .. &                     .. &               $<$0.73 &                             .. \\
       He I 3187 &                  3188.6660 &            $>$-0.41 &                            .. &                     .. &               $<$0.61 &                             .. \\
      Ti II 3239 &                  3239.9712 &            $>$-0.34 &                            .. &                     .. &               $<$0.50 &                             .. \\
     Ne III 3342 &                  3343.1420 &           $>$-0.61 &                            .. &                     .. &               $<$0.93 &                             .. \\
      S III 3721 &                  3722.6870 &           $>$-0.35 &                            .. &                     .. &               $<$0.48 &                             .. \\
   {[}O II] 3727 &                  3727.0920 &             -20.6 &                          0.26 &                 146. &                68.8 &                           0.85 \\
   {[}O II] 3729 &                  3729.9000 &             -25.1 &                          0.25 &                 164. &                83.5 &                           0.83 \\
            H$\eta$ &                  3836.4790 &              -0.62 &                          0.25 &                   3.04 &                  2.22 &                           0.89 \\
 {[}Ne III] 3869 &                  3869.8610 &              -3.97 &                          0.64 &                   8.07 &                 14.2 &                           2.28 \\
           H$\zeta$ &                  3890.1580 &              -3.57 &                          0.81 &                   9.11 &                 12.8 &                           2.91 \\
\hline
\end{tabular}
\caption{Keck ESI line flux measurements for \twentythreelong .  The reported numbers are measured from the summed spectra of the four images.  Images 2, 3 and 4 are 0.91, 1.09, and 0.40 times as bright, respectively, as image 1. flux$_{\mathrm{obs}}$ and $\delta$ flux$_{\mathrm{obs}}$ denote             the observed flux and uncertainty respectively. W$_\mathrm{r,fit}$ and $\delta$ W$_{\mathrm{r,fit}}$ denote the rest-frame             equivalent width measured and the corresponding uncertainty in \AA\, respectively, with negative values indicating emission (rather than absorption) lines. For cases of             non-detection (i.e. $<$3 $\sigma$ detection), the 3$\sigma$ upper limit on equivalent widths and fluxes are quoted.}
\end{table*}

\seventeen\ does not appear to be substantially reddened at any location.  Region C is consistent with an \ebv\ of zero, though the $1\sigma$ measurement uncertainties in the H$_{\alpha}$ and H$_{\beta}$ lines allow \ebv to be as high as 0.07.  The region with the most reddening is region B, with E(B-V) $=$ 0.10 $\pm$ 0.05.  Regions A and D fall in between these values, with reddenings of 0.02 $\pm$ 0.05 and 0.04 $\pm$ 0.09 respectively, though like region C, they are both consistent with zero.  There is little evidence of spatial variation in reddening in \seventeen\ except that region B may be slightly more reddened than the other regions as it is the only region inconsistent with zero reddening at about the 2$\sigma$ level.

\twentythree\ shows signs of strong variation in the Balmer line ratios.  Regions A and B are more reddened than any of the regions in \seventeen , having E(B-V) $=$ 0.39 $\pm$ 0.07 and 0.38 $\pm$ 0.06 respectively. Region C, however, is notably less reddened, with E(B-V) $=$ 0.06 $\pm$ 0.13, which is consistent with zero and more comparable to \seventeen\ than to the other regions within \twentythree. The H$\alpha$/H$\beta$ ratio for region C differs from those of regions A and B by about 2.8$\sigma$.  Of the three regions within \twentythree , region C appears the bluest in broadband imaging, which is consistent with this apparent lower level of extinction, but could also be due, for example, to lower metallicities or a younger stellar population.

In principle, reddening can also be determined using the H$\beta$/H$\gamma$ ratio.  In practice, the smaller wavelength separation relative to H$\alpha$/H$\beta$ and the relative faintness of H$\gamma$ make it less useful than H$\alpha$/H$\beta$.  In addition, at the low spectral resolution of the G102 grism, H$\gamma$ is blended with [O~III]~4363\AA, which is responsible for the high uncertainties in our quoted H$\gamma$ flux.  We find that, in these spectra, the uncertainties on the H$\gamma$ flux prevent a meaningful measurement of the reddening via H$\beta$/H$\gamma$.

\subsubsection{Ratios Sensitive to Ionization Parameter and Metallicity: O32 and R23}\label{sec:results-o32_r23}
Several diagnostic strong line ratios are sensitive to both ionization parameter and metallicity to varying degrees.  Those observable in the wavelength range in the grism spectroscopy for \seventeen\ and \twentythree\ include the following:
\begin{itemize}
\item {\textbf{R23}} $\equiv$ ([O III] 4959, 5007~\AA\  + [O II] 3727, 3729~\AA) / H$\beta$
\item {\textbf{O32}} $\equiv$ [O III] 4959, 5007~\AA\ / [O II] 3727, 3729~\AA 
\item {\textbf{Ne3O2}} $\equiv$  [Ne III] 3869~\AA\ / [O II ] 3727, 3729~\AA
\item {\textbf{O3H$\beta$}} $\equiv$ [O III] 5007~\AA\ / H$\beta$
\end{itemize}
It should be noted that some of these ratios have varying definitions in the literature.  For example, \citet{Levesque:2014hq} use only the [O~II] 3727~\AA\ line in the definition of Ne3O2, but for this paper we prefer to include the 3729~\AA\ line because the blended sum is all that can be measured in spatially resolved regions due to the low spectral resolution of the \HST\ G102 grism.  Similarly, R23 is sometimes defined without the 3729~\AA\ line, as in \citet{Nakajima:2013ir} (who, incidentally, also choose to parameterize the [O~III] to [O~II] ratio as [O~III] 5007~\AA\ / [O~II] 3727~\AA ), though others (\citealt{Onodera:2016aa} for example) include the 3729~\AA\ line.

Of these ratios, O32, Ne3O2, and O3H$\beta$ are primarily sensitive to ionization parameter, while R23 is primarily sensitive to metallicity.  O3H$\beta$ has a strong dependence on pressure, a quantity for which we do not have spatially--resolved indicators, and as a result is a poor indicator of either ionization parameter or metallicity.  For this reason, we do not include this diagnostic in our analysis.

\begin{figure}[b]
\includegraphics[width=3.4in]{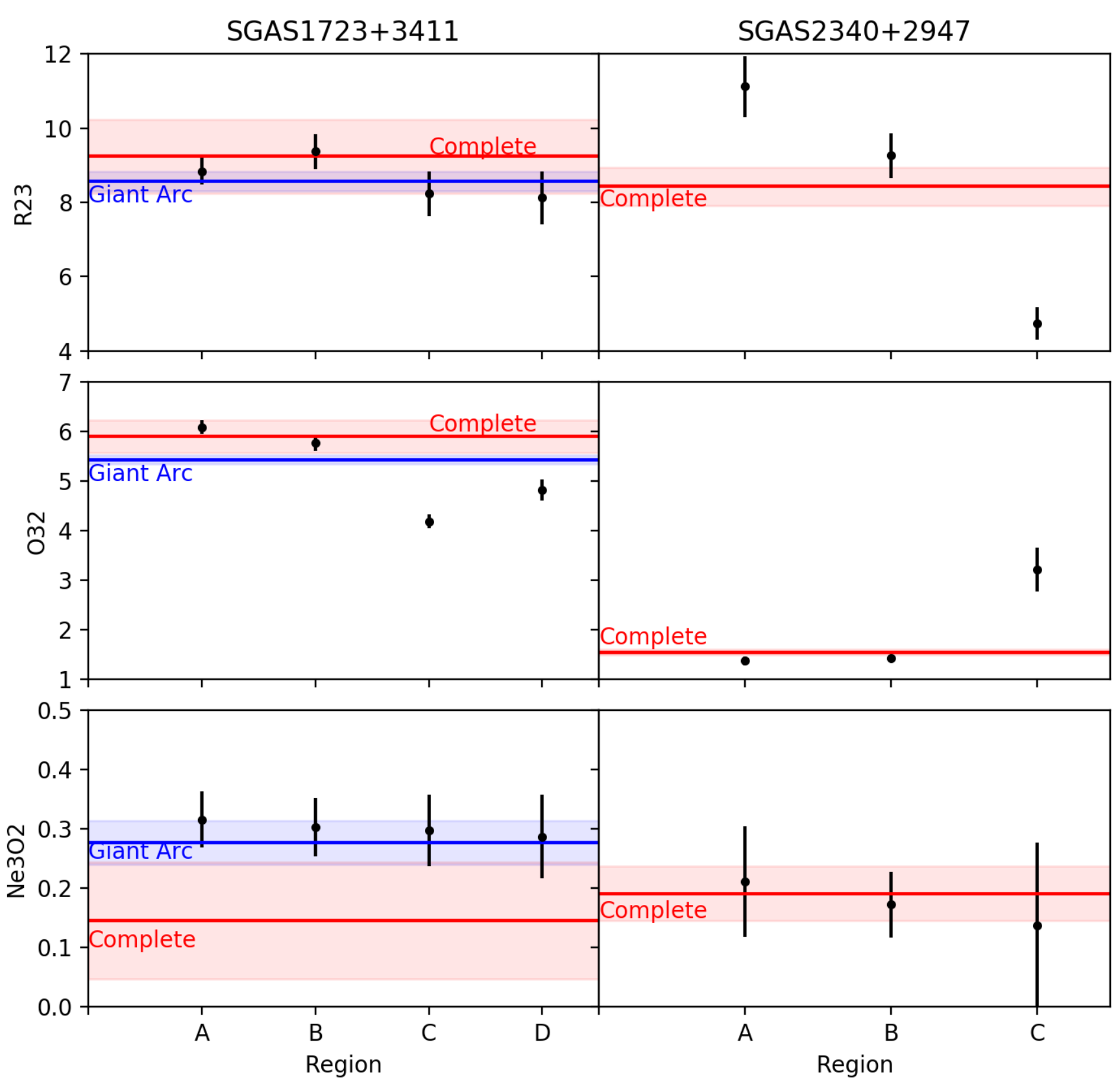}
\caption{Comparison of the R23, O32, and Ne3O2 values for spatially resolved regions (black points) and the values measured from spatially integrated spectra of either complete images (red) or the \seventeen\ giant arc (blue).  There is a suggestion of spatial variation of R23 and clear variation of O32 in \seventeen\ and as well as significant variation in both R23 and O32 in \twentythree .  It is also clear from the disagreement between the spatially integrated values and some of the spatially resolved values that integrated spectra do not provide a complete picture of the physical conditions inside galaxies like these (see section~\ref{sec:disc-ORN}).}
\label{fig:allRatiosWBands}
\end{figure}

The top row of Fig.~\ref{fig:allRatiosWBands} shows the spatial variation in the diagnostic R23, which is primarily sensitive to metallicity.  The values for the regions of \seventeen\ are somewhat tightly bunched around 8.5; the largest and smallest R23 values---those of regions B and D---disagree at only about the 1.5$\sigma$ level.  By contrast, there is substantial variation across the sub-regions of \twentythree.  As we saw with the reddening diagnostics, region C is quite different from the other two regions.

The middle and bottom rows of Fig.~\ref{fig:allRatiosWBands} show the spatial variation in the diagnostics O32 and Ne302, which are primarily sensitive to ionization parameter.  The regions in \seventeen\ show generally higher O32 than those in \twentythree, which suggests that the ionization parameter in \seventeen\ is systematically higher than in \twentythree.  Both sources show significant spatial variation in O32.  In \twentythree, once again, region C stands out, differing from A and B at the 4$\sigma$ level, while A and B are in close agreement with each other.  In \seventeen , regions A and B show significantly higher O32 values than regions C and D (at the 5--10$\sigma$ level depending on the pair of regions); regions A and B agree with each other within 1.5$\sigma$ while there is 2.5$\sigma$ difference between C and D. 

To summarize, within \seventeen , regions A and B have higher O32 values than regions C and D; the same pattern is seen (at lower significance) for the Balmer decrement and R23.  Within \twentythree , region C displays higher O32, lower R23, and less reddening than the rest of the source.  
 
As for Ne3O2, the values are slightly higher in \seventeen\ than in \twentythree .  We find no evidence of spatial variation of Ne3O2 in either galaxy.  Because of the relatively faint [Ne~III] line involved and its potential blends with nearby lines, the fractional uncertainties in Ne3O2 are relatively large (up to 25\% in \seventeen\ and larger for \twentythree), and greatly exceed any apparent variation in the observed Ne3O2 values.

\subsubsection{Star Formation Rate}
So far, the values that we have considered are all magnification independent because they rely only on relative fluxes, and gravitational lensing is achromatic.  We can also look for variation in the H$\alpha$ flux, which is a well-known indicator of star formation \citep{Kennicutt:1998ki}.  Unlike the previous values discussed in this paper, though, this quantity is dependent on the magnification of each region.  Table 4 shows the magnifications of each region determined using the lens models discussed in section~\ref{sec:lensmodels}.

\begin{figure}[]
\includegraphics[width=3.4in]{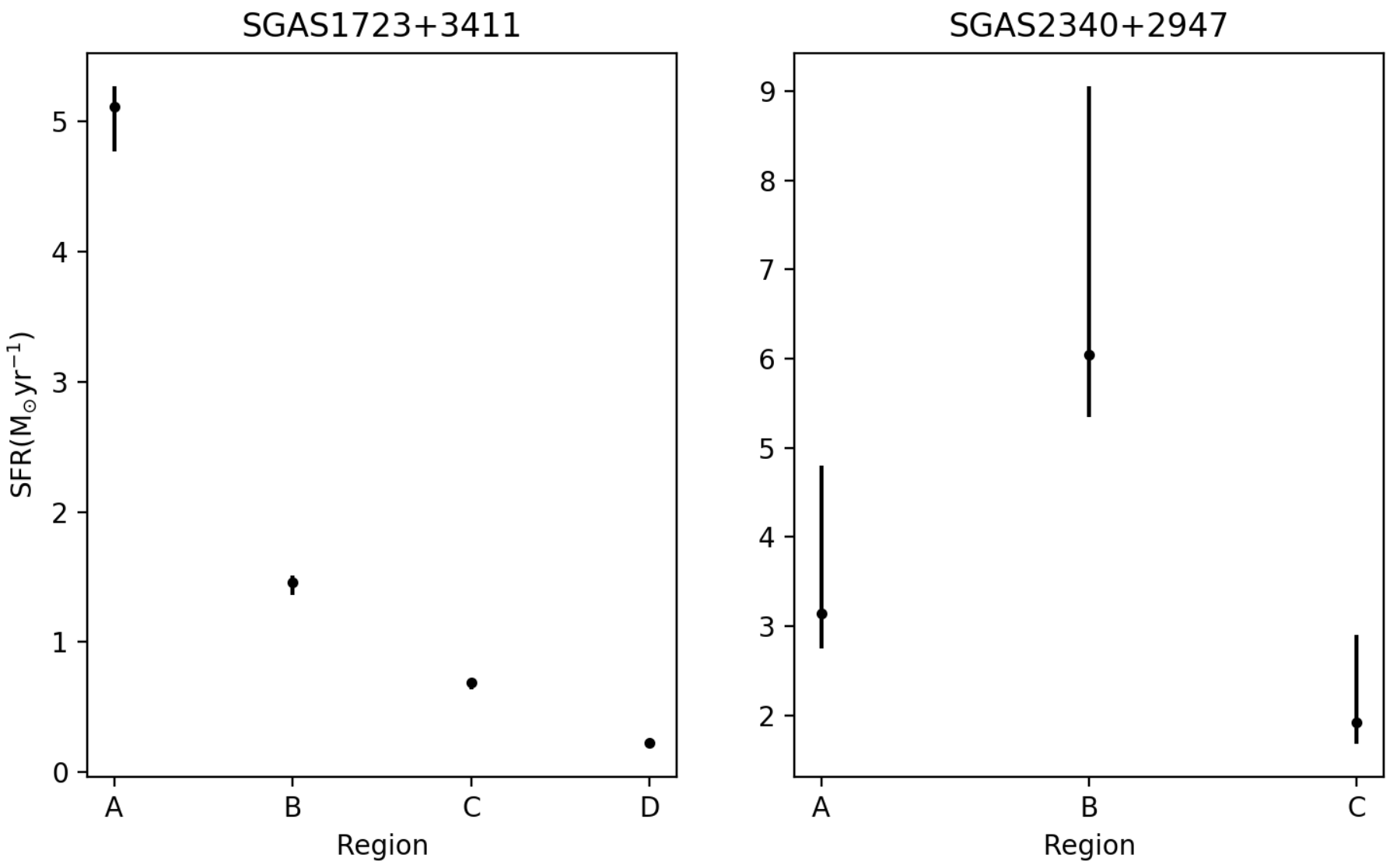}
\caption{Variation in star formation rate from region to region within \seventeen\ (left) and \twentythree\ (right).  Both show evidence of gradients.  \twentythree\ shows a negative radial gradient, with regions A and C forming stars less rapidly than region B. The SFR of \seventeen\ is driven primarily by clump A.}
\label{fig:SFR}
\end{figure}

For each region in each source, we applied the conversion from H$\alpha$ to SFR from \citet{Kennicutt:1998ki} using the spatially-resolved reddening corrections and magnification corrections.  There is significant spatial variation in the star formation rates across the two galaxies, as shown in Fig.~\ref{fig:SFR}.  \twentythree\ is forming stars quite rapidly. The central region (B), exhibits the most star formation, about $6.0_{-0.7}^{+3.0} M_{\odot} yr^{-1}$, while the regions near the edges of the source (A and C) are forming about  $3.1_{-0.4}^{+1.7} M_{\odot} yr^{-1}$ and  $1.9_{-0.2}^{+1.0} M_{\odot} yr^{-1}$ of stars respectively.  This suggests a negative radial gradient in star formation rates, but also suggests asymmetric star formation.  As a check of consistency, the SFR determined by stacking all 4 complete images is $13.2_{-1.7}^{+7.1}$ \Msol $yr^{-1}$, which agrees with the sum of the 3 subregions.  Significant uncertainties remain, however, due to the lack of extra constraints for the lens model.

\seventeen\ is also undergoing rapid star formation driven by two of the three apparent clumps.  Regions A and B are forming $5.11_{-0.34}^{+0.16} M_{\odot}$ and $1.46_{-0.1}^{+0.05} M_{\odot}$ per year, respectively, while regions C and D are forming stars at much lower rates, ($0.69_{-0.05}^{+0.02} M_{\odot} yr^{-1}$ and $0.23_{-0.01}^{+0.01} M_{\odot} yr^{-1}$).  The northern complete image of \seventeen\ (image 3) appears to have a SFR of $7.91_{-0.39}^{+0.41} M_{\odot} yr^{-1}$, similar to the sum of these regions, suggesting that the portion of the galaxy visible in the arc accounts for nearly all of the star formation in \seventeen .  The low surface brightness diffuse light at the outskirts of \seventeen\, then, likely contains relatively little star formation (about 0.4 \Msol yr$^{-1}$).  Since the bright clumps are concentrated near the center of this galaxy, \seventeen\ may also have a negative gradient in the star formation rate.  However, it is clear that the star formation is patchy and asymmetric, dominated by only two clumps.

As a check of consistency, the SFR derived from fitting the integrated spectrum of the entire giant arc (i.e., summing the spectrum of the whole arc and fitting the lines) is $6.68_{-0.20}^{+0.44} M_{\odot} yr^{-1}$, slightly lower than what we measure by summing the measured line fluxes of the individual regions (i.e., extracting spectra for each region and fitting the lines in each), but in agreement within about 1$\sigma$.

Although \seventeen\ shows evidence of patchy star formation dominated by two clumps, those clumps are centrally-located relative to the faint whisps of light evident at the ends of the arc and in image 3 that extend much further out from the center of the galaxy.  Both sources, therefore, exhibit centrally-concentrated star formation.  Such excesses have been interpreted as ``inside-out star formation" in other work (e.g., \citealp{Nelson:2016ds} at similar redshifts, and \citealp{ellison2018} at lower redshifts).

\subsection{Ionization Parameters and Metallicities}\label{sec:results-logUandZ}
In \S~\ref{sec:results-o32_r23} we showed that the observable line ratios O32 and R23 vary in a statistically significant way.  We now consider whether this can robustly be attributed to real differences in physical parameters.   To address this question, we must convert between the observables and physical parameters by referring to photoionization models, in this case MAPPINGS v5.1 models (described in the following subsection) to infer the physical parameters logU (ionization parameter) and Z (metallicity).

\subsubsection{MAPPINGS Photoizonization Model Grids}
We use calibrations from 
\citet{kewley:2019aa}, and \citet{kewley:2019ab} (henceforth K19a and K19b, respectively) and \citet{nicholls:2020}, which are based on the latest version of results from the MAPPINGS v5.1 photoionization code \citep[see][]{Sutherland:1993aa, Dopita:2013bj}.  MAPPINGS v5.1 includes the local Milky Way region elemental abundances for 30 elements. Nebular elemental abundances for sub-solar metallicites are scaled according to \citet{Nicholls:2017du}. For depletion of nebular elements on to dust grains, the K19a,b models adopt the parametric models of \citet{Jenkins:2009aa} with a Fe depletion value of -1.5 dex. K19a,b use the atomic data for the 30 elements from the CHIANTI 8 database \citep{DelZanna:2015ie}. The MAPPINGS photoionization code self-consistently computes the ionization structure of the nebulae, accounting for dust absorption, radiation pressure, grain charging, and photoelectric heating of small grains \citep{Groves:2004aa}.

The ionization parameter is the ratio of the local Lyman photon flux (cm$^{-2}$s$^{-1}$) to the local hydrogen density (cm$^{-3}$). This ionization parameter ($q$), with units of velocity, can be related to a dimension-less ionization parameter ($U$), which we use in this paper, via the speed of light: $U = q/c$. MAPPINGS defines the ionization parameter at the inner edge of the nebula. The photoionization, excitation and recombination is calculated in a detailed, self-consistent manner in linear increments of 0.02 step size through the nebula. See \citet{LopezSanchez:2012aa} for a full description of the models and geometries.

The K19a,b calibrations use constant pressure models with plane parallel geometry. The ISM pressure values range from $\log{(P/k)} = 4.0$ to $\log{(P/k)} = 9.0$ in increments of 0.5 dex. This pressure corresponds to the total mechanical energy flux imparted on the nebula by the driving stellar source, through contributions from both stellar winds and supernovae. These models compute a detailed temperate and density structure throughout the nebulae, dictated by the metallicity and ionization structure.

For the purpose of calibrating electron density diagnostics, K19a use constant density models with electron densities ranging from $\log{(n_e/cm^3)}$ = 0 to $\log{(n_e/cm^3)} = 5$ in increments of 0.5 dex. These models compute a temperature structure in the nebula, but unlike the isobaric models do not allow for a density structure. K19a point out that the isobaric models are the most realistic given that the sound crossing timescale in typical nebulae is shorter than the cooling/heating timescale which allows for the pressure to equalize throughout the nebula.

The metallicity ($12 + \log{\mathrm{O/H}}$) values in the models are constrained by the stellar tracks, which use a coarse grid of $12 + \log{\mathrm{O/H}}$ = 7.63, 8.23, 8.53, 8.93 and 9.23. The models are computed for a range of ionization parameter values from $\log{(q)} = 6.5$ to $\log{(q)} = 8.5$, in increments of 0.25 dex. This range corresponds to $\log{(U)} \simeq -4$ to $\log{(U)} \simeq -2$, which covers the typically observed values in H~II regions \citep{Dopita:2000aa}.

\subsubsection{Variation in ionization parameter and metallicity based on MAPPINGS models}\label{sec:gridstuff}
The grids used in this section are derived from models where the pressure, $\log(P/k)$, is 6.0 for \seventeen\ and 6.5 for \twentythree , as determined by the [O~II]3727/3729 ratio measured from the ESI spectra of the giant arc (consisting of two nearly complete images) in \seventeen\ and the sum of the integrated spectra of the complete images of \twentythree .

\begin{figure*}[]
\centering
\includegraphics[width=7in]{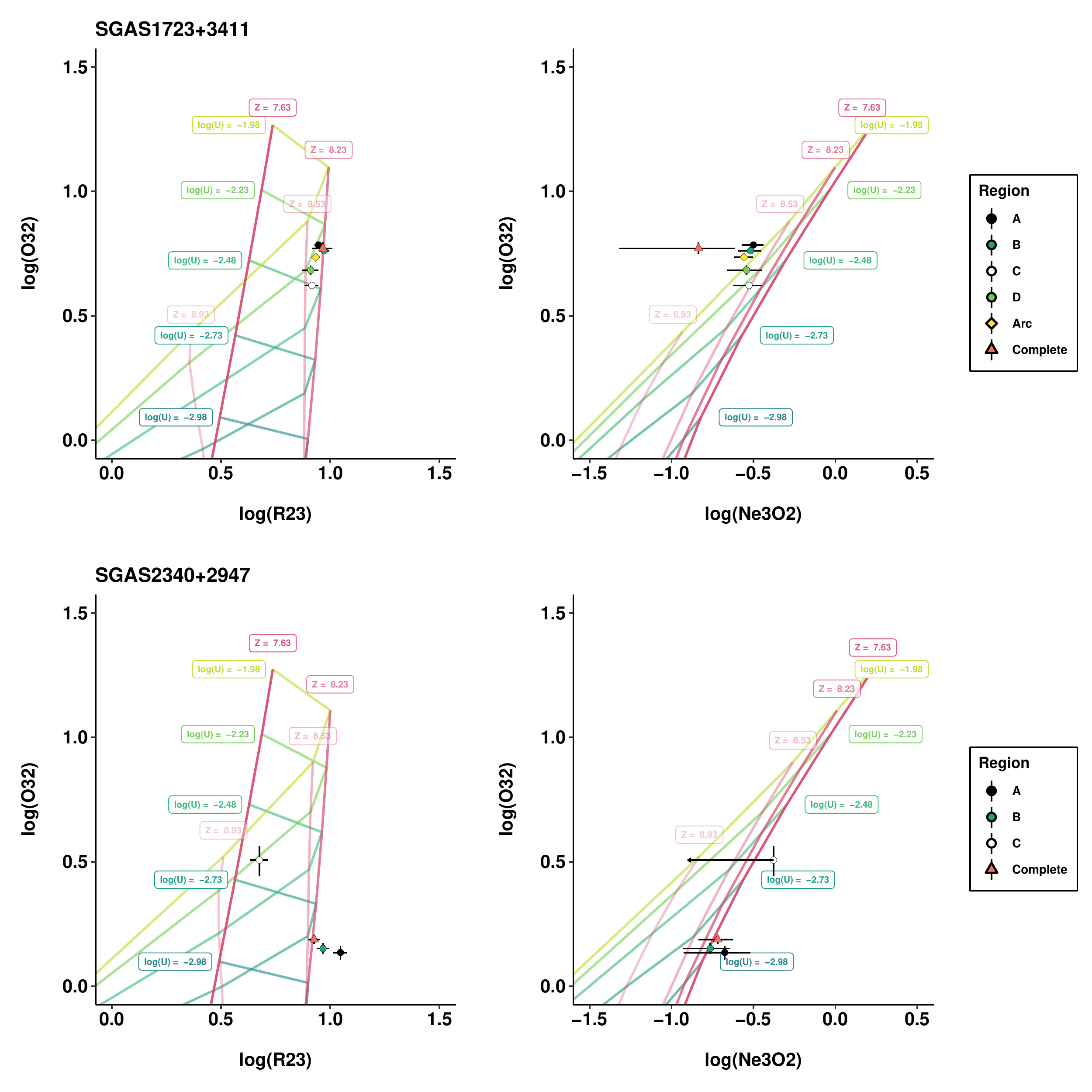}
\caption{Spatial variation in physical conditions logU and Z (12 + logO/H) as determined by the O32-R23 and O32-Ne3O2 planes and MAPPINGS models.  Red lines indicate curves of constant metallicity, with the color saturation scaling with metallicity.  Other colors correspond to curves of constant ionization parameter.  Note that the MAPPINGS grid in the O32-R23 plane is double-valued and therefore folds over on itself.  Spatial variation in these sources appears to be primarily driven by ionization parameter (and the O32 observable).  However, the R23 values of \twentythree also indicate significant variation in metallicity.}
\label{fig:UandZ}
\end{figure*}

In Fig.~\ref{fig:UandZ}, we plot the observed strong line ratios for each region in \seventeen\ and \twentythree , as well as for the northern complete image and the giant arc in \seventeen\ and the stacked spectra of the complete images of \twentythree\ (three images from one roll, and one image from the other as summarized in Table 2) in the O32--R23 and O32--Ne3O2 planes.  These points are overlaid on the MAPPINGS grids,  from which we infer ionization parameter ($\log U = \log q/c$) and metallicity ($Z =  12 + \log(O/H)$).  Characteristically, the grids in the O32--R23 plane are double-valued; most of our points fall near the double-valued region or where the grids fold over.  Consequently, the interpretation of these inferred parameters will have some amount of degeneracy in that there is both a high-metallicity interpretation and a low-metallicity interpretation.  The [N~II]/H$\alpha$ flux ratio from GNIRS, approximately 6.2\% for region A of \seventeen\ and 8.4\% region B of \twentythree , suggest that these regions do fall almost exactly on the fold. The MAPPINGS models suggest a metallicity near $12+\log(O/H) \sim 8.23$ for both sources, though it could be as high as 8.6 for \seventeen\ if it is on the higher metallicity branch. Similarly, the first-order formula of \citet{Pettini:2004bq} returns a metallicity of $8.21 \pm 0.04$ for \seventeen\ and $8.29 \pm 0.09$ for \twentythree .  Their third-order formula returns $8.19 \pm 0.03$ for \seventeen\ and $8.25 \pm 0.07$ for \twentythree .  Additionally, the ratio of [O~II] 2470\AA\ to [O~III] 3727/3729\AA\ can be used as a temperature diagnostic, which predicts a temperature of 11,384$^{+3067}_{-2759}$K for \twentythree (similar to, but slightly higher than the what Rigby \etal, in prep, finds for \seventeen\ ), and also suggests a moderate metallicity \citep{nicholls:2020}.

For \seventeen , we see that most of the variation is in O32, not in R23.  The model grids indicate that this largely corresponds to variation in the ionization parameter.  Regardless of which metallicity branch is assumed, the $\log U$ in each region is between $-2.23$ and $-2.48$, for a spread of a little less than 0.25 dex in ionization parameter.  If the points are on the low-metallicity branch, then they fall near, but slightly below $Z = 8.23$ with very little spread.  However, if they are on the high metallicity branch, they fill the region $Z = 8.23$ to $Z = 8.53$, with a spread of about 0.30 dex.

The O32--Ne3O2 plane suggests that \seventeen\ has a slightly higher ionization parameter than the O32--R23 plane implies, with values falling between  $\log U = -1.98$ and $\log U=-2.23$; the spread is similar, still about 0.25 dex.  Again, the variation is almost entirely in the ionization parameter and is most apparent in the O32 dimension. Unlike the O32--R23 grid, the Ne3O2--R23 grid is not double-valued, and the Ne3O2 values suggest that the metallicity in every region of \seventeen\ is somewhat higher than we inferred from the O32--R23 grids.  Still, if we assume that this means that \seventeen\ falls on the high-metallicity branch in O32--R23 space, then those metallicities agree with the Ne302--R23 metallicities.

\twentythree\ is quite different from \seventeen\ in a number of ways.  Of particular interest is the more extreme spatial variation.  While regions A and B are very similar in both ionization and metallicity, region C is nothing like them.  Both grids, O32 vs R23 and O32 vs Ne3O2, show that regions A and B, as well as the complete image, have ionization parameters that are tightly bunched about halfway between $\log U=-2.73$ and $\log U=-2.98$ (i.e., around $\log U = -2.85$).  This is not the case for region C, which has a much higher ionization parameter.  While the lack of a robust detection in Ne3O2 diminishes its utility as an indicator of ionization parameter, the location of region C in the O32--R23 plane suggests that it has an ionization parameter of either $\log U \sim -2.65$ (low metallicity branch) or $-2.23$ (high metallicity branch).  In the low metallicity case, the offset between region C
and the other regions in ionization parameter is about the same size as the spread in ionization parameters across the different regions of \seventeen .  In the high metallicity case, the offset may be more than twice as large.  Unlike \seventeen , though, this variation is driven by a single unique region---the rest of \twentythree\ is nearly uniform in $\log U$.

The interpretation of metallicity for \twentythree\ is a little more difficult because some of the points, particularly in the O32--R23 plane fall outside the model grid.  This is likely a real phenomenon, as these points lie in a region of the O32--R32 plane that is populated by z$\sim$2 galaxies in MOSDEF \citep{Reddy:2018aa}, z$\sim$2--3 galaxies in \citet{Nakajima:2013ir}, and z$\sim$3.3 galaxies in \citet{Onodera:2016aa}.  Why the models do not account for this is unclear, but there are several sources of uncertainty in the photoionization models that could lead to this.  Nonuniform conditions within emission regions, uncertainties in the calibrations of strong line diagnostics, and uncertainties in atomic data all contribute to uncertainties in the location of grid points.  If we assume that the points that fall off of the O32--R23 grids for \twentythree\ are actually at or near the grid's fold, then the metallicity is around $Z=8.23$. 

Small uncertainties in the Ne3O2 ratio correspond to large uncertainties in the inferred $Z$ from the O32--Ne3O2 grid.
Still, that grid indicates that regions A and B and the complete image fall at relatively low $Z$, around $7.63$ to $8.23$; however, it does not rule out metallicities as high as $8.53$.  Region C, based on O32 vs R23, can either be around $Z \sim 7.7$ if on the low metallicity branch, or $\sim 8.8$ if on the high metallicity branch.  Since region C has a higher ionization parameter than the other regions and is substantially less reddened, it would seem reasonable for it to be on the low metallicity branch.  However, we do not have a [N~II]/H$\alpha$ measurement for this region to break the degeneracy.

Overall, these two galaxies exhibit variation in ionization parameter of at least $0.25$ dex.  The metallicity variation is probably smaller, but the double-valued O32--R23 grid makes it difficult to say definitively.  The variations in ionization and metallicity in \seventeen\ manifest as slight variation across the different regions.  Regions A and B are more similar to each other in all  of $\log U$, $Z$, $E(B-V)$, and SFR, than they are to regions C or D (which themselves are more like each other than they are like A or B).  Because of the proximity of A to B and C to D, this looks like the physical conditions are, perhaps, spatially correlated, but they are not necessarily representative of a radial gradient (since the morphology is so clumpy and the variations are driven by differences in the clumps) and may simply be indicative of asymmetry.  It is much clearer, though, that the variation in \twentythree\ is driven almost entirely by one single spatial region (C) that is very different from the others, especially in $\log U$, $Z$, $E(B-V)$.  This, too, appears more like an asymmetry than a radial gradient except in SFR.  It is possible that these asymmetries are indicative of a recent or ongoing merger (e.g., region C being accreted by a galaxy composed of regions A and B).

\section{Discussion}\label{sec:discussion}

In the previous sections, we presented observational evidence for significant spatial variation in strong line ratios and the physical conditions that drive them across two galaxies at $z \sim 1.4$.  This finding has significant implications for interpreting current observations of field galaxies, as well as implications for planning future observations; here, we explore some of these potential impacts.  We focus primarily on observable strong line ratios, rather than physical parameters like $U$ and $Z$, because direct observables are model-independent.

\subsection{Reddening}\label{sec:reddening_var_imp}
The finding that reddening can vary spatially across $z\sim 1.4$ galaxies should not be surprising, but it importantly affects the interpretation of other observed strong line ratios.  
Fig.~\ref{fig:reddeningWBands} shows the H$\alpha$/H$\beta$ ratios for each spatially-resolved region, compared to the ratios from the spatially-integrated spectra.  While there may be slight tension between the value for region B in \seventeen\ and the value derived from the spectrum of the whole arc, there is relatively little variation in reddening throughout the regions visible in the giant arc, and no conflict with the line ratio measured from the spectrum of the complete image, despite the fact that it contains more of the underlying source galaxy than just the regions contained in the arc.  This is not particularly surprising, though, since the part of the source that is not imaged in the arc is relatively small, with low surface brightness compared to the rest of the source.  It is actually somewhat remarkable that the reddening in the individual regions agrees so well with the reddening of the arc and the complete image despite the spectra of the latter two being essentially wavelength-by-wavelength flux-weighted averages of the spectral properties of regions A--D.  This did not have to be the case, as shown clearly by the H$\alpha$/H$\beta$ ratios in \twentythree.  The line ratio in the spectra of the complete images is much more representative of the bright, similarly-reddened regions A and B than it is of the much fainter and much less reddened region C.

\begin{figure}[]
\includegraphics[width=3.4in]{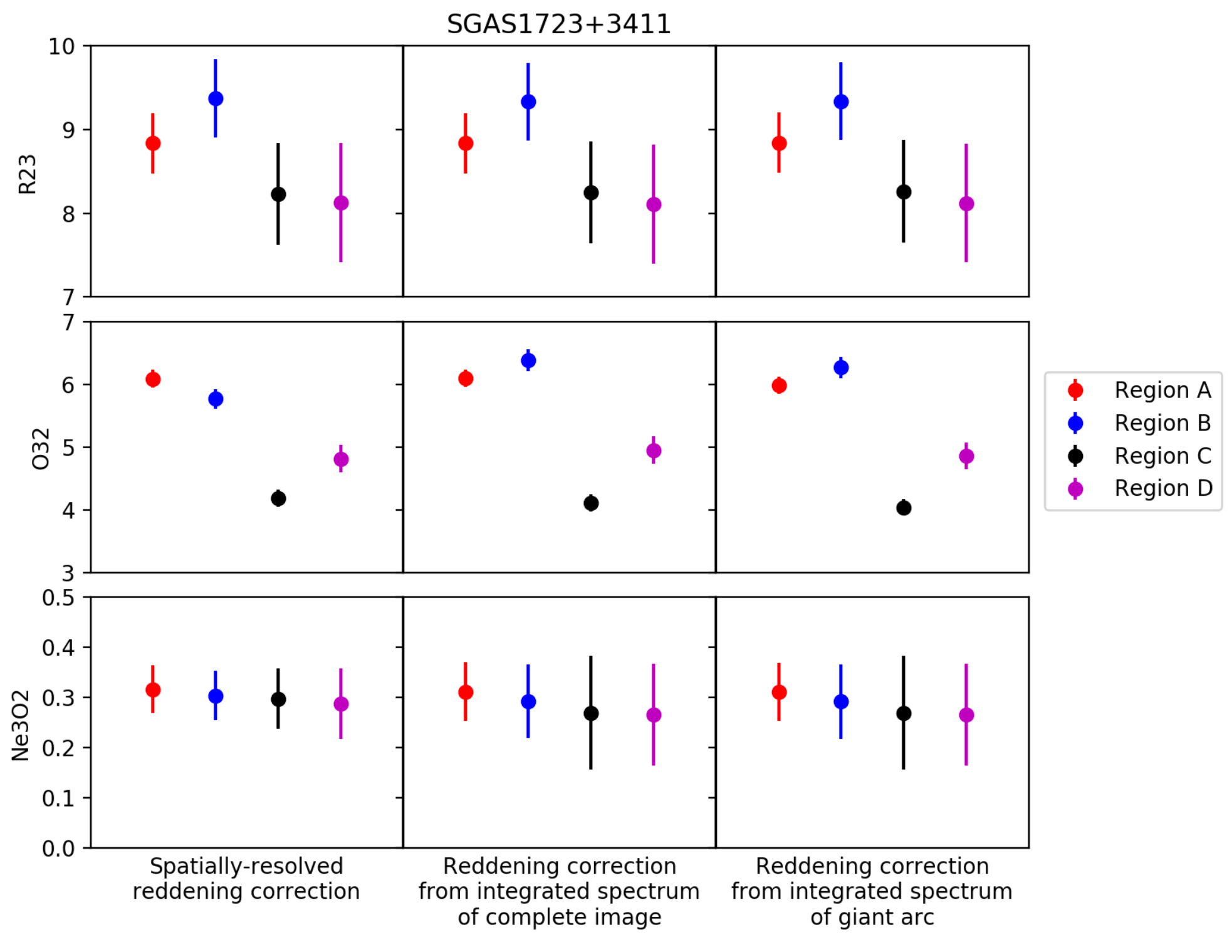}
\caption{From top to bottom, R23, O32 and Ne3O2 of individual spatially-resolved regions in \seventeen\ color-coded by region.  Each column of plots displays these values determined using different reddening corrections.  Using a reddening correction derived from the  spatially-resolved spectra(left column) clearly influences the O32 value of region B and brings it below the value of region A, compared to using reddening corrections from the spectrum of the complete image (middle column), or the spectrum of the entire giant arc (right column).  This demonstrates the potential importance of obtaining spatially-resolved reddening indicators rather than assuming a single uniform reddening for an entire galaxy.
}
\label{fig:reddeningImp1723}
\end{figure}

\begin{figure}[]
\centering
\includegraphics[width=3.4in]{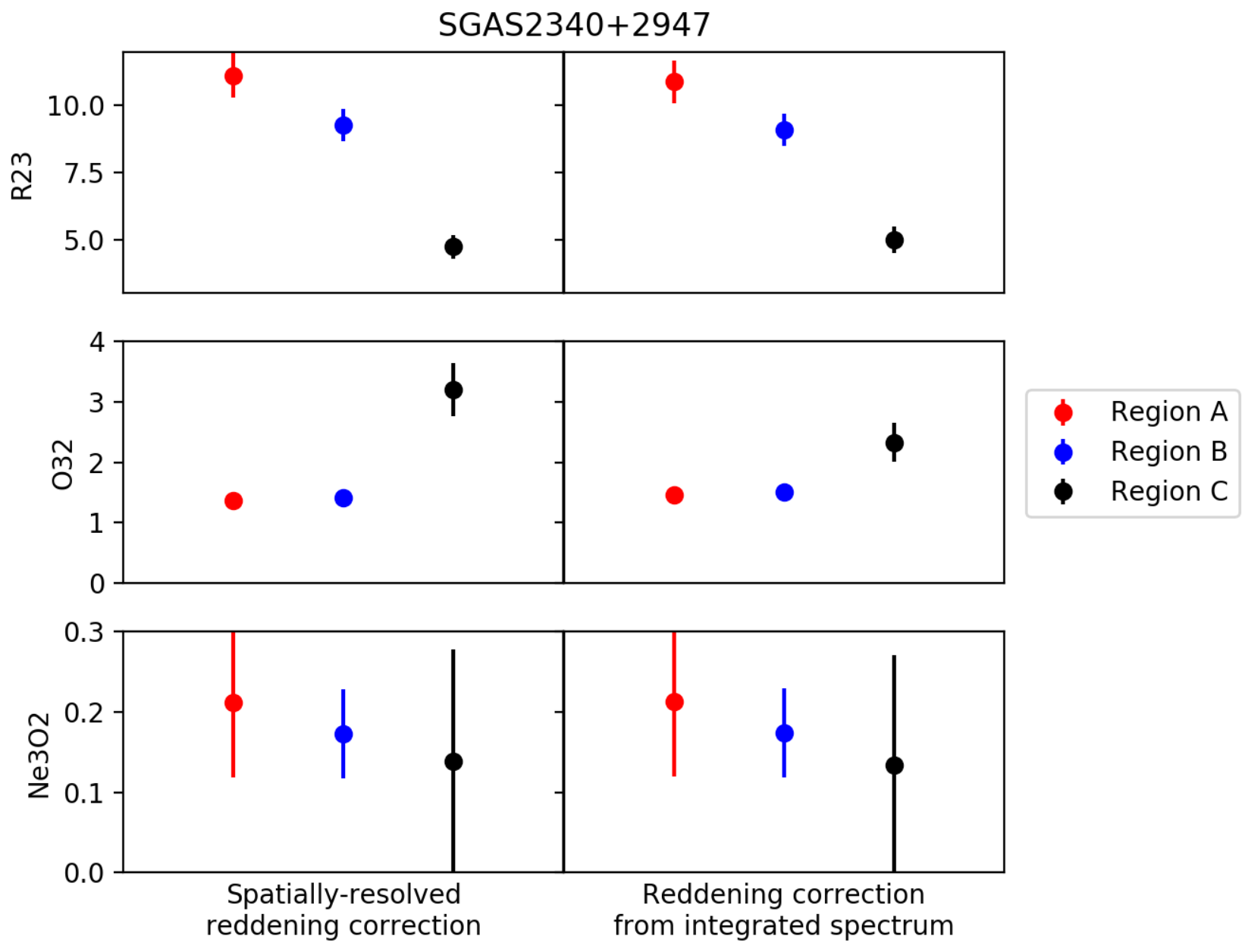}
\caption{The same as Fig.~\ref{fig:reddeningImp1723} except for \twentythree.  In this case there is no giant arc, so the left column uses spatially-resolved H$\alpha$ to H$\beta$ ratios to de-redden the spectra, while the right column uses the ratio from the stack of the complete images.  Once again this has a noticeable impact on O32.  In particular, O32 is higher in region C when spatially-resolved reddening corrections are used.  This is likely because its reddening is much lower than that what would be inferred from the spectrum of the complete image as shown in Fig.~\ref{fig:reddeningWBands}.}
\label{fig:reddeningImp2340}
\end{figure}

With this in mind we explore the consequences of using spatially-integrated reddening corrections to spatially-resolved spectra, as one might want to do for higher redshift targets where the H$\alpha$ line is shifted redward of the G141 grism.  Figs.~\ref{fig:reddeningImp1723} and \ref{fig:reddeningImp2340} show the difference between the measured values of R23, O32 and Ne3O2 with a spatially-resolved reddening correction applied compared to what we would get using a global correction from either the complete images of \seventeen\ or \twentythree , or the giant arc for \seventeen .   Fig.~\ref{fig:reddeningImp1723} shows the impact on the reddening-corrected R23, O32, and Ne3O2 ratios.  While individual values do not typically change by much, the slight excess in reddening observed in region B is enough to have an important effect on its O32 ratio.  By not using spatially-resolved values of H$\alpha$/H$\beta$, we would incorrectly infer that the value of O32 in region B is actually higher than in region A, even though this is not the case.  While the overall change is low, this type of uncertainty could influence searches for spatial gradients in ionization parameter or metallicity.   In \twentythree\ (Fig.~\ref{fig:reddeningImp2340}), we see no such changes in ordering, but the value of O32 in region C is noticeably underestimated.  Since high O32 correlates with leakage of ionizing photons, searches for LyC leakers that depend on indirect indicators like O32 could miss candidates if they apply spatially integrated reddening corrections.

The fact that spatially-resolved reddening corrections can have this kind of impact is a particularly important consideration for planning observations in the near-term because there is only a small window of redshifts where the [O~II] 3727/3729 doublet and the H$\alpha$ line are both observable with space-based, spatially-resolved spectroscopy at this point in time (z$\sim$1.15--1.58 with the \HST\ WFC3/IR grisms).  Without both of these features one cannot determine a reddening-corrected value of the O32 diagnostic, for instance.  

These two objects were chosen to fall within that window, but most potentially interesting targets will not.  It may be tempting to extend this window by assuming spatially uniform reddening and using a spatially-integrated H$\alpha$ measurement from ground-based spectroscopy, for example, to determine the reddening.  Doing so would expand the observable redshift range to $z \sim 2.4$ by making the reddest line of interest the [O~III] 5007 doublet instead of H$\alpha$.  While this may work for sources like \seventeen\ where the reddening is relatively uniform spatially, it could cause problems for sources like \twentythree\ where there is significant variation in reddening.  In fact, as we have seen here, even the relatively low variation in reddening within \seventeen\ may be enough to meaningfully affect the interpretation of strong line ratios of a region like B.

When reddening cannot be measured in a spatially resolved way, it is likely that regions of higher extinction will be missing or under-represented.  This presumably also affects \HST\ surveys that use broad-band rest-frame UV light as a tracer of star formation as well as H$\alpha$--based surveys with the \HST\ grisms.  Real progress will come with spectroscopy from the IFUs on \jwst 's NIRSPec and MIRI instruments, which will be able to measure spatially-resolved H$\alpha$/H$\beta$ ratios---and better yet, Paschen $\alpha$ / H$\alpha$ ratios---over a much larger range of redshift than is possible with the \HST\ WFC3-IR grisms.  Until then, interpretation of spatially-resolved spectroscopic studies of galaxies with redshifts above 1.58 with the \HST\ grisms will suffer from serious uncertainties due to the unconstrained reddening variation.

\subsection{O32, R23, Ne3O2}\label{sec:disc-ORN}
We do not detect statistically significant spatial variation in Ne3O2.  This is at least partly due to the fact that [Ne~III]~3869 is difficult to disentangle from nearby lines at the low spectral resolution of the \HST\ grisms.  In addition, in the MAPPINGS models, while Ne3O2 is more sensitive to $Z$ than to $\log U$, small uncertainties in Ne3O2 can translate to large uncertainties in $Z$, such that it does not usefully constrain $Z$.  R23, which does show significant spatial variation, more powerfully discriminates between different Z values than does Ne3O2.  The downside is that the MAPPINGS grid is double-valued in the O32--R23 plane, and the spectrographs with high spatial resolution at these wavelengths (WFC3-IR grisms) lack the spectral resolution to obtain other spatially--resolved indicators to break the degeneracy.  Ultimately, high spectral resolution, high spatial resolution spectroscopy of lensed galaxies is necessary to break the degeneracy by obtaining, for instance, a spatially-resolved [N~II]/H$\alpha$ ratio.  This will be firmly within the capabilities of \JWST\  IFU spectroscopy, for instance (and indeed, \seventeen\ will be observed as part of the JWST Early Release Science program, TEMPLATES; PI: Rigby).

We confidently detect spatial variation in the O32 and R23 ratios.  By necessity, most of the spectroscopy of distant ($z\ga 1$) galaxies is spatially integrated, so we now compare the spatially-integrated values of O32, R23 and Ne3O2 to the spatially-resolved values to see what, if anything, we are missing in typical observations.  Fig.~\ref{fig:allRatiosWBands} shows this comparison.  There are clear differences in the results. The value of R23 measured from the complete image of \twentythree\ is most in agreement with the value for region B.  However, the values for regions A and C lie significantly higher and lower, respectively, than the value for the complete image.  Meanwhile, even though the O32 value for the \twentythree\ complete images is near the values for regions A and B, it is significantly below the value for region C.  

\seventeen\ further illustrates how spatially integrated spectra can obscure the actual physical conditions within a galaxy.  While the O32 value for the complete image agrees with the values in regions A and B, it disagrees with the values for regions C and D, and is only barely compatible with the value for the integrated spectrum of the entire giant arc.  The value for the giant arc, interestingly, lies below the values for regions A and B, but above the values for regions C and D.  Even though there is general agreement between the spatially-resolved and the spatially-integrated values of R23, it is clear that, in general, the physical conditions in subregions are not necessarily well represented by the values inferred from spatially integrated spectra of either complete images or giant arcs.  Just as studies of the stacked spectra of many galaxies miss the unique characteristics of individual galaxies, so too do the integrated spectra of single galaxies miss the unique characteristics of individual physical regions within them.

\begin{figure}[b]
\includegraphics[width=3.4in]{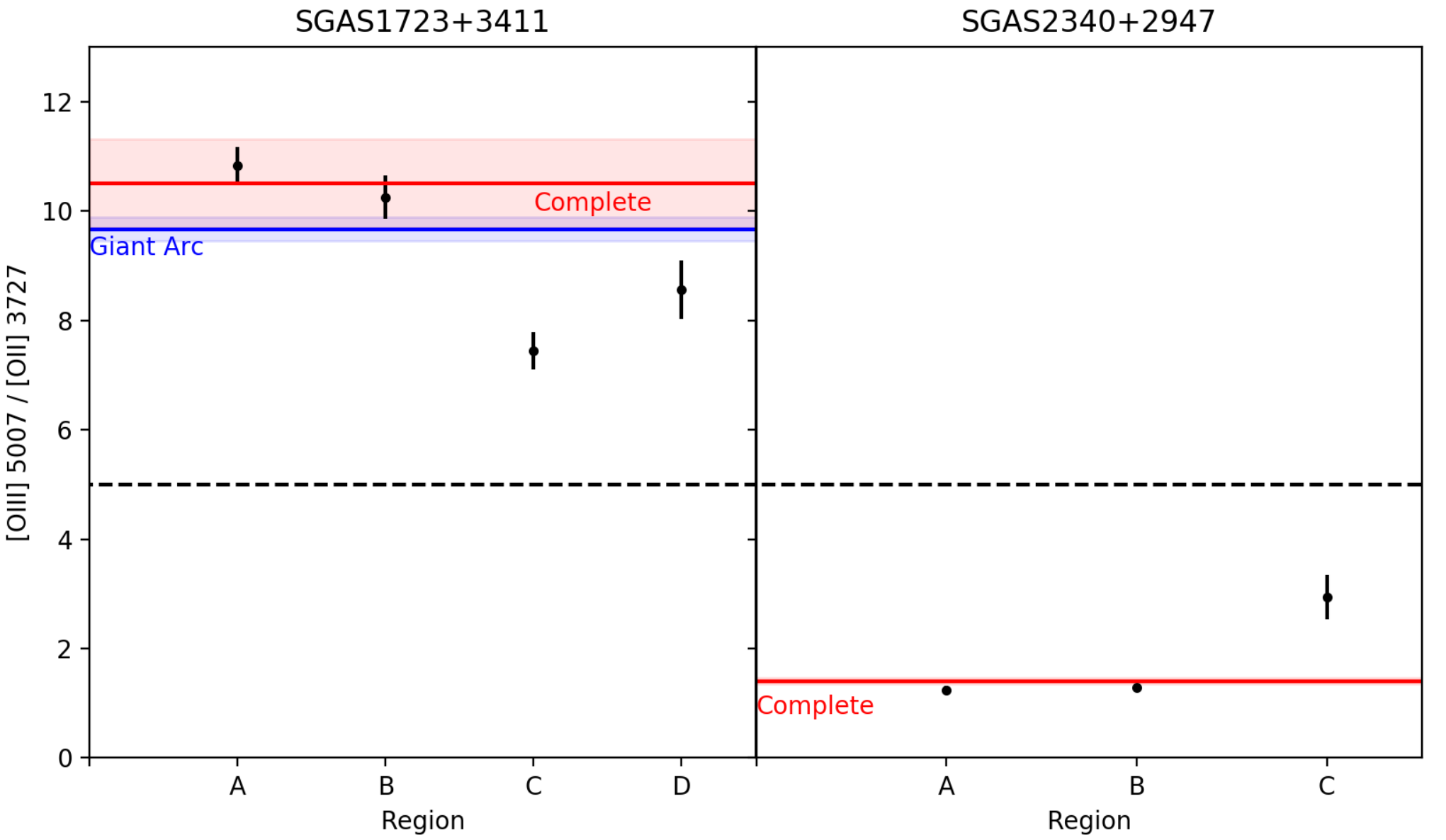}
\caption{A slightly different parameterization of the [O III] to [O II] ratio, this time [O III] 5007 / [O II] 3727.  This ratio has been suggested to be an indicator of possible LyC leakage if it exceeds 5 (dashed line).  Since the spatially-resolved grism spectra cannot spectrally resolve the 3727/3729 doublet, the value of that ratio is set based on the ground-based spatially-integrated ESI spectra (though based on the Figs.~\ref{fig:reddeningWBands} and ~\ref{fig:allRatiosWBands}, it is not necessarily good to assume that ratios like this do not vary spatially).  While the spatially resolved spectra do not result in identifications of LyC leakers where the spatially integrated spectra did not for these two specific sources, \twentythree\ in particular suggests that this may not always be the case.}
\label{fig:oiiilyc}
\end{figure}

\subsection{Implications for finding galaxies that leak ionizing photons}
Galaxies with high O32 ratios are found to be leaking ionizing photons at much higher rates than the general galaxy population (e.g.\  \citealt{Izotov:2018aa, Nakajima:2013ir}).  In particular, \citet{Izotov:2018aa} suggests that a flux ratio of [O~III] 5007 / [O~II] 3727 $\ga 5$ is necessary but not sufficient for a galaxy to be leaking ionizing photons.  (Note: While this is a ratio of [O~III] to [O~II], it is not the same definition that we have been using for O32 in this paper.  See discussion in \S\ref{sec:results-o32_r23}.)  In Fig.~\ref{fig:oiiilyc}, we plot this ratio (using their definition) for each of the regions in \seventeen\ and \twentythree\ as well as for the spatially-integrated spectra to determine how using spatially-integrated spectra would have influenced a search for LyC leakers that included \seventeen\ and \twentythree.  The [O II] 3727 and 3729 lines are not resolved in the grism spectra, so in order to assign a value of [OII ] 3727 to each region, the blended [O II] 3727 + 3729 values were multiplied by the spatially-integrated 3727/3729 ratio determined from the ground-based ESI data.  For \seventeen , this is the value for the integrated spectrum of the giant arc.  For \twentythree, this is the value for the summed integrated spectra of all four complete images.

In neither case would the complete or giant arc values have suggested that the source was not a leaker when small regions inside it were leaking.  For \seventeen , all four regions lie well above the threshold of 5.  The giant arc and complete image also yield large values.  However, the values for regions C and D are somewhat lower than for regions A and B, and they drag down the value measured for the giant arc.  So while \seventeen\ would likely have remained on any list of LyC candidates based on this criterion alone, it may have been given a lower priority for follow-up than it deserves, since region A is a stronger candidate for leakage than the spectrum of the sum of the giant arc would suggest.

\twentythree, while having no region that exceeds [O III] 5007 / [O II] 3727 $=$ 5, still suggests that it is quite possible to miss a candidate by looking only at spatially-unresolved spectra.  The value for region C is much higher than what is implied by the complete image.  And while it is still well below 5, it is not hard to imagine a similar case where the points are all shifted slightly upward and the complete image shows low [O III] 5007 / [O II] 3727, but one small spatial region, like region C, lies above the threshold.  If the physical processes that lead to the conditions that enable the escape of ionizing photons are spatially isolated events rather than large galaxy-wide phenomena, then surveys using spatially-integrated spectra can be expected to systematically miss candidates with only one or a few leaking regions.  In fact, it is already known that at least some objects exist where LyC leakage is a local, rather than galaxy-scale event (e.g., \citealp{RiveraThorsen:2019vx}), so this concern must be taken seriously as a potential source of bias.

\section{Conclusions}
To summarize, using spatially-resolved grism spectroscopy of two highly magnified star-forming galaxies at $z = 1.33$ and $z = 1.42$, we have explored the extent of spatial variation in observable strong diagnostic lines and the model-dependent physical parameters that they imply.  We have also examined potential implications of such variation for interpreting existing data and for planning future observations, including with the \jwst\ IFUs.  The following bullet points summarize our main findings.
\begin{itemize}
\item {There is statistically significant variation in strong-line ratios like O32 and R23, but any variation in the Ne3O2 diagnostic is below the level that we can measure in this type of data.}
\item {These variations lead to spreads of about 0.2-0.5 dex in metallicity and ionization parameter, depending on which metallicity branch these regions reside on in the R23--O32 plane.}
\item {The Balmer decrement, H$\alpha$/H$\beta$ varies significantly within a single source galaxy.  Spatially-integrated spectra yield values of this ratio (and of R23 and O32) that may differ substantially from that of individual, distinct physical regions within a single galaxy.  Furthermore, applying spatially-integrated reddening corrections to spatially-resolved line fluxes can improperly influence the inferred values of other diagnostic ratios like R23 and O32.}
\item{The star formation in these sources is concentrated near the center of these galaxies, suggesting that they are perhaps undergoing inside-out star formation.  Each source also displays a rather large degree of asymmetry in star-formation rates.  For \seventeen\ this is apparent even among the bright, star-forming clumps concentrated near the center of the source.  The morphology of its star formation is perhaps more characterized by its clumpiness than a radial gradient.  It is worth considering whether calculating azimuthally-averaged gradients in properties like SFR is still an effective metric when applied to high-resolution data since asymmetries in annuli would make such a measurement quite misleading in a source like \seventeen . \citet{Curti:2020} raise the same issue in the context of metallicity maps.  Perhaps, in high spatial-resolution data, a metric like the clumpiness indices or Gini coefficients of properties like SFR, metallicity, etc. could be more informative.}
\item{If O32 is used as an indirect indicator of possible LyC leakage, then sources in which only a single or a few small regions are leaking LyC photons may systematically be undercounted in surveys that use spatially integrated spectra. Extreme O32 ratios in a single region, for example, can be drowned out by less extreme ratios in the rest of the source, causing a more moderate O32 value to be measured from the integrated spectrum.}
\item {\seventeen , based on the O32 values in each of its spatial regions, is a candidate for LyC leakage.  \twentythree\ has less extreme ratios and is not as good of a candidate.}
\end{itemize}



\acknowledgments
{This research was supported by an appointment to the NASA Postdoctoral Program at the Goddard Space Flight Center, administered by Universities Space Research Association through a contract with NASA.

Based on observations made with the NASA/ESA Hubble Space Telescope, obtained from the Data Archive at the Space Telescope Science Institute, which is operated by the Association of Universities for Research in Astronomy, Inc., under NASA contract NAS 5-26555. These observations are associated with program \# 14230.

Support for program 14230 was provided by NASA through a grant from the Space Telescope Science Institute, which is operated by the Association of Universities for Research in Astronomy, Inc., under NASA contract NAS 5-26555.

Some of the data presented herein were obtained at the W.M. Keck Observatory, which is operated as a scientific partnership among the California Institute of Technology, the University of California and the National Aeronautics and Space Administration. The Observatory was made possible by the generous financial support of the W.M. Keck Foundation.  We acknowledge the very significant cultural role and reverence that the summit of Maunakea has always had within the indigenous Hawaiian community. We are most fortunate to have the opportunity to conduct observations from this sacred mountain.

Some of the observations reported here were obtained at the MMT Observatory, a joint facility of the University of Arizona and the Smithsonian Institution. 

Some of the data presented herein were obtained at the Gemini Observatory, which is operated by the Association of Universities for Research in Astronomy, Inc., under a cooperative agreement with the NSF on behalf of the Gemini partnership: the National Science Foundation (United States), the National Research Council (Canada), CONICYT (Chile), Ministerio de Ciencia, Tecnología e Innovación Productiva (Argentina), and Ministério da Ciência, Tecnologia e Inovação (Brazil).

We would also like to thank Glenn Kacprzack for his advice and assistance in the Keck ESI data reduction process.}

\appendix
\section{Stellar mass estimates}
Total stellar masses for each of these objects were determined using the MCMC-based stellar population synthesis and parameter inference code, Prospector (that is based on the python-FSPS framework, with the MILES stellar spectral library and the MIST set of isochrones; \citealp{conroy:2010fsps};  \citealp{emceehammer}; \citealp{prospector}; \citealp{leja:2017a}).

As input to this code, we used the demagnified magnitudes of each of these objects from the avaiable HST and Spitzer bands.  For \seventeenlong , we used the magnitudes of the giant arc, which required us to apply a correction to account for the fact that it is a merging set of two nearly (but not totally) complete images.  For \twentythreelong the photometry was summed across the three most magnified images, ignoring the less magnified counterimage (image 4) that is closer to the BCG and more contaminated by light from cluster galaxies.  For each source, fluxes were corrected for Milky Way reddening.

\begin{figure*}[]
\centering
\includegraphics[width=7in]{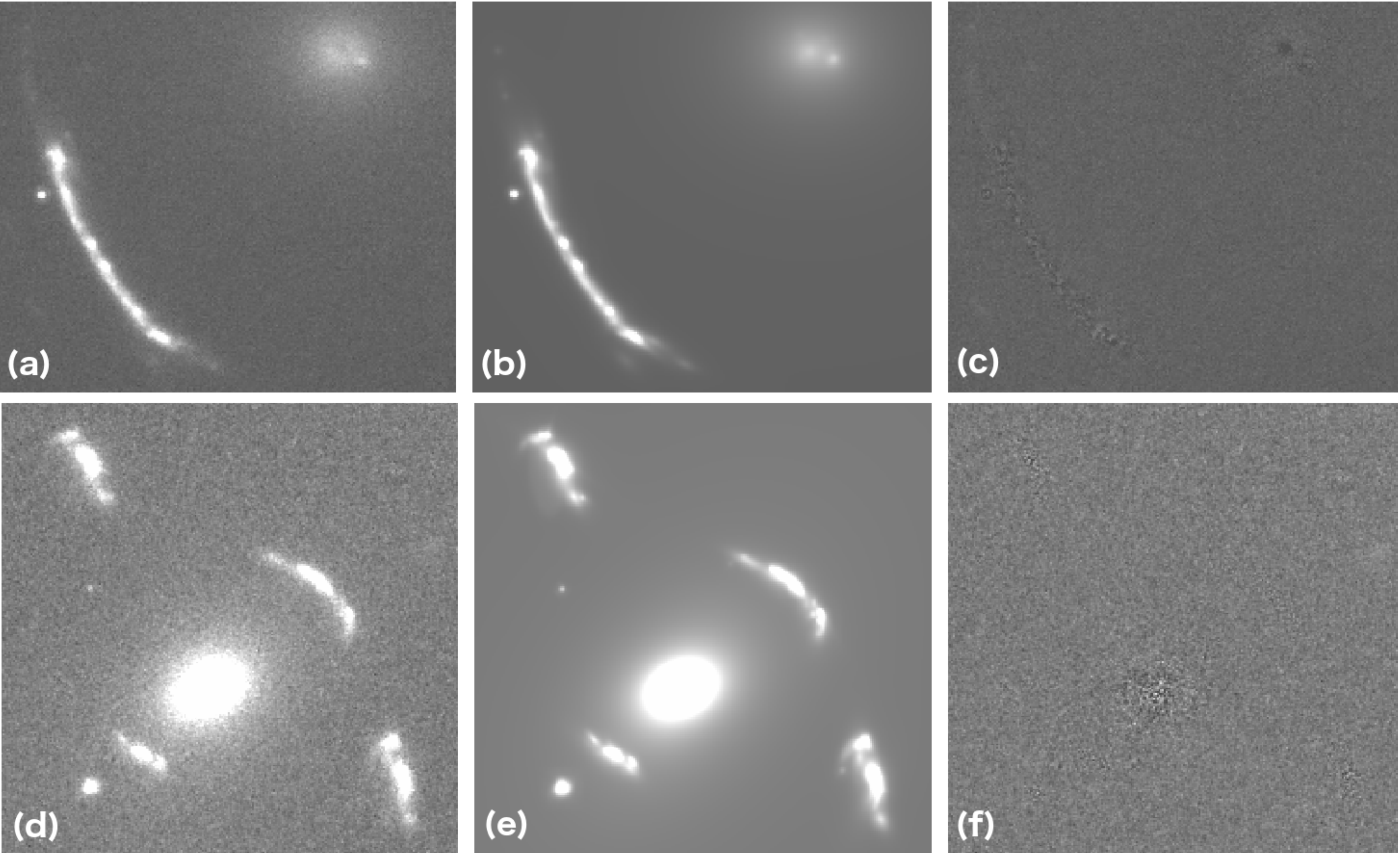}
\caption{Direct images, {\tt{GALFIT}} models, and residuals for \seventeen\ and \twentythree .  (a): The stack of the F390W and F775W images of \seventeen .  (b): The {\tt{GALFIT}} model of (a).  (c): The difference (residual) of (a) and (b). (d): The F814W image of \twentythree .  (e): The {\tt{GALFIT}} model of (d). (f): The difference (residual) of (d) and (e). }
\label{fig:galfit}
\end{figure*}

We created detailed models for the light profiles of each of these objects using {\tt{GALFIT}} \citep{peng:2002} and calculated the photometry from these models.  For the HST images, the models describe the lensed galaxies and the BCG, as well as any other cluster members or nearby foreground or background galaxies that may contaminate the light profiles of the lensed sources.  Point spread functions (PSFs) were derived from the data directly, by summing appropriately isolated point source images surrounding the model region.  The initial model was constructed using a sum of the F390W and F814W (or F775W) images.  This initial model was then propagated to individual HST filter images by re-optimizing it, allowing for small ($\sim20\%$) variations in fitted structural parameters in each step in wavelength away from the initial model, in addition to freely varying the magnitudes of the model components.  Examples of these models are shown in Fig.~\ref{fig:galfit}  Total magnitudes of each lensed source were then derived by summing the flux across all relevant structural components that comprise each image. Simple aperture photometry - derived from HST images with the fitted central lens galaxy image removed, is consistent with these more complex measurements.

The Spitzer PSF has broad enough wings that this it was necessary to include additional model components to describe other nearby objects to enable a robust sky measurement in the modeling.  The HST-derived model of the lensed sources is, in practice, more complex than necessary to describe them in the Spitzer data. In fits in which the source components are essentially unconstrained, it is thus typical that some components trend toward zero flux. Experiment shows that constrained fits, in which the components are all required to be significant, produce the same measurement in that the total flux summed across all the components that describe a given lensed image of the source is consistent.

Demagnified and Milky Way reddening-corrected magnitudes are tabulated for the \seventeen\ and \twentythree\ in Table 6.  These magnitudes serve as constraints for the stellar population modeling.  In these models for \twentythree\, we assumed a parametric star formation history (simple tau model, with e-folding time $\tau$ and star formation start time $sf_{start}$, both in Gyr), dust attenuation applied to all light from the galaxy
(in units of opacity at 5500\AA ), metallicity $log(Z/Z_{\odot})$ (where $Z_{\odot}=0.0142$), 
and total mass formed in the galaxy (in \Msol), as free parameters. The dust extinction and metallicity have priors covering the 2$\sigma$ range suggested by the available spectroscopy.  For \seventeen\, which was consistent with zero reddening, we used a dust-free model.  These models each assumed a Kroupa IMF (Kroupa 2001) and that nebular continuum 
and line emission are present.  We assumed the WMAP9 cosmology \citep{hinshaw:2013aa} where necessary.

\begin{figure*}[h]
\centering
\includegraphics[width=7in]{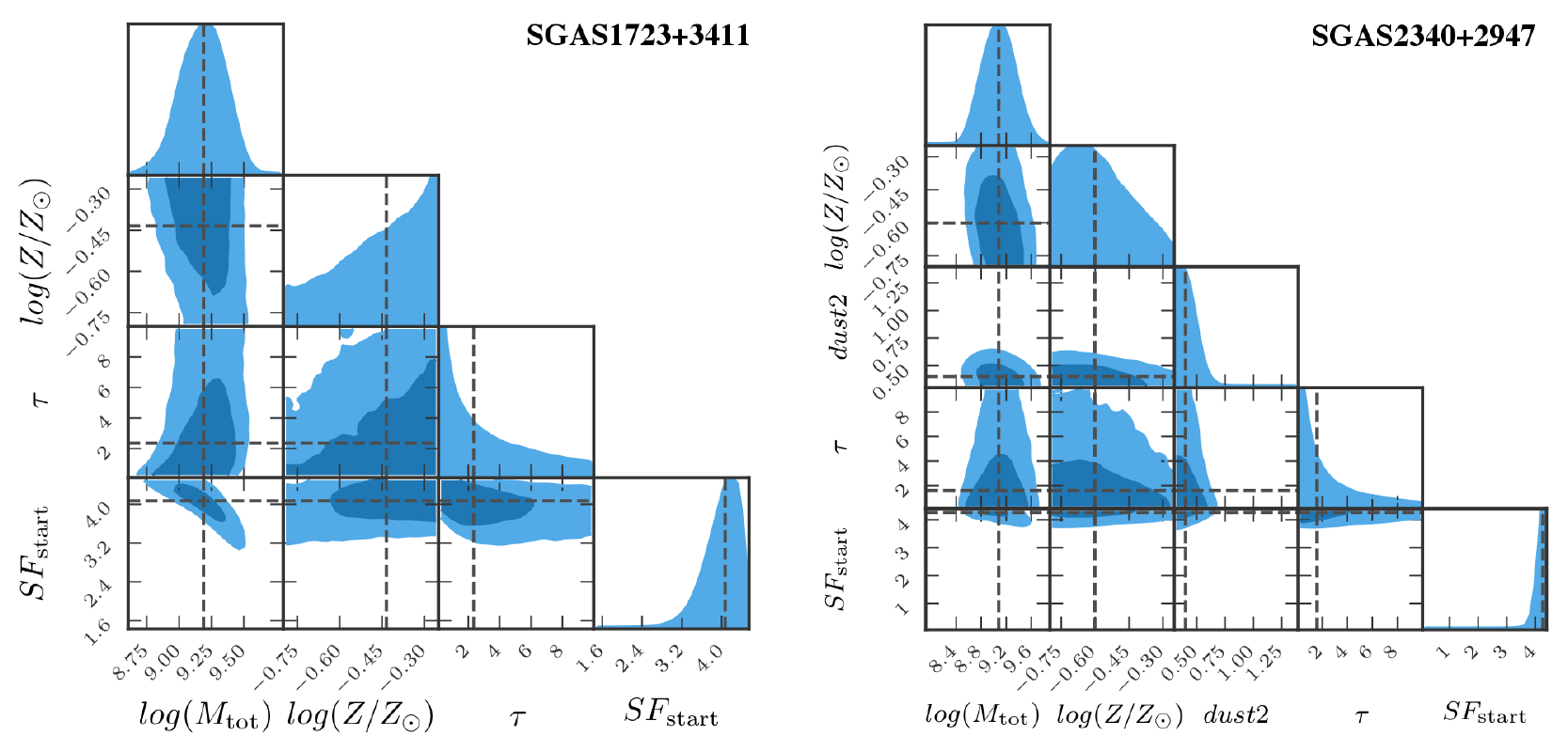}
\caption{Corner plots for the SED models of \seventeen\ (left) and \twentythree\ (right).  M$_{tot}$ is the stellar mass in units of solar masses.  Z/Z$_{\odot}$ is the metallicity relative to solar metallicity.  $\tau$ is the star formation e-folding time in Gyr.  SF$_{start}$ is the time, in Gyr, at which star formation began.  \seventeen\ was modeled as if it were dust-free (consistent with the Balmer decrement), while our model for \twentythree\ included fitting for dust within a range determined by the measured Balmer decrement and its uncertainties.  The parameter, dust2, is the optical depth of the dust at 5500\AA\ .}
\label{fig:cornerPlots}
\end{figure*}

\begin{figure*}[]
\centering
\includegraphics[width=6in]{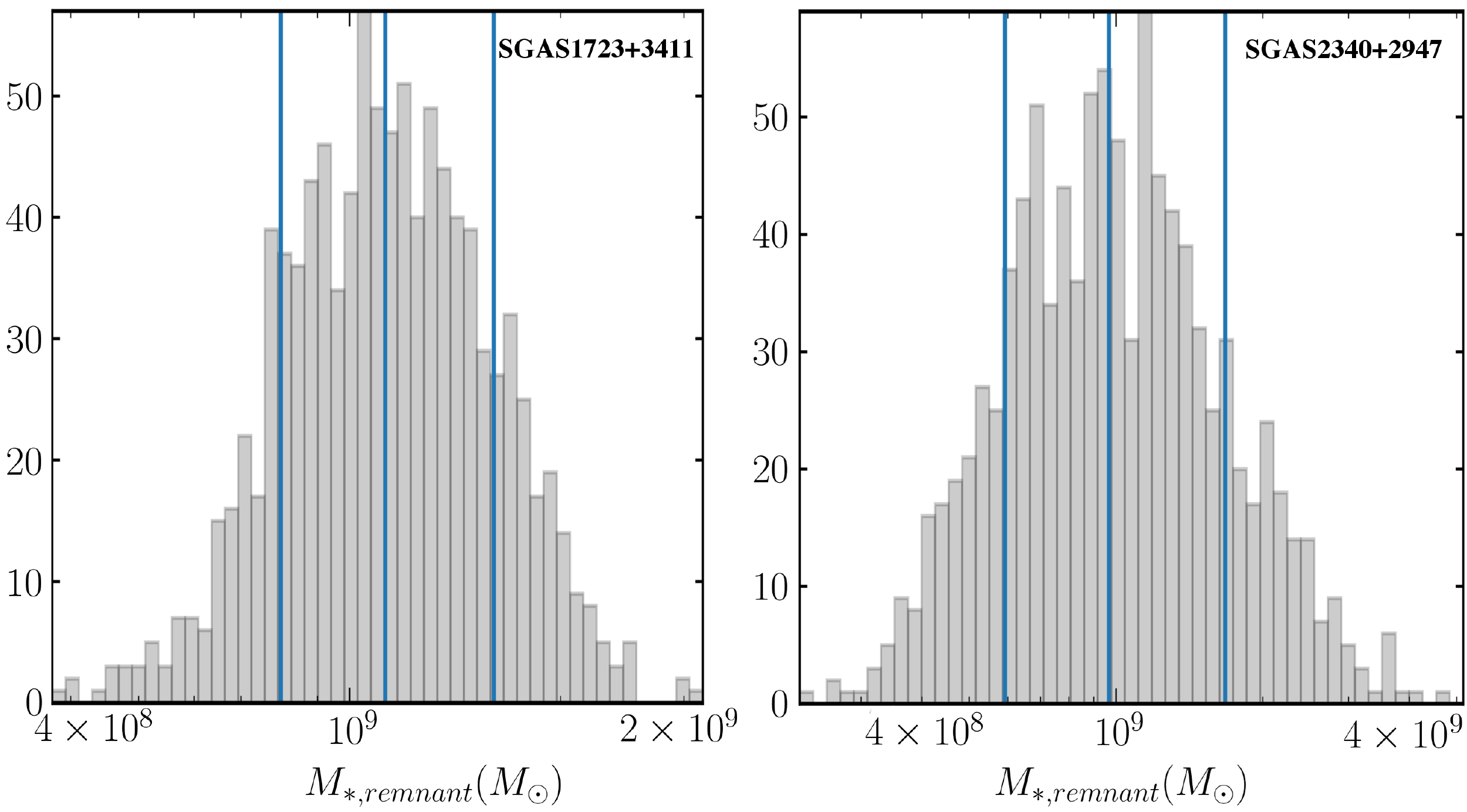}
\caption{Posterior distribution for the observed remnant stellar masses, constructed from SPS models corresponding to 10000 random chains for \seventeen\ (top) and \twentythree\ (bottom).  Vertical lines, from left to right, denote the $16^{th}$, $50^{th}$, and $84^{th}$ percentiles.}
\label{fig:massHists}
\end{figure*}

\begin{figure*}[]
\centering
\includegraphics[width=6in]{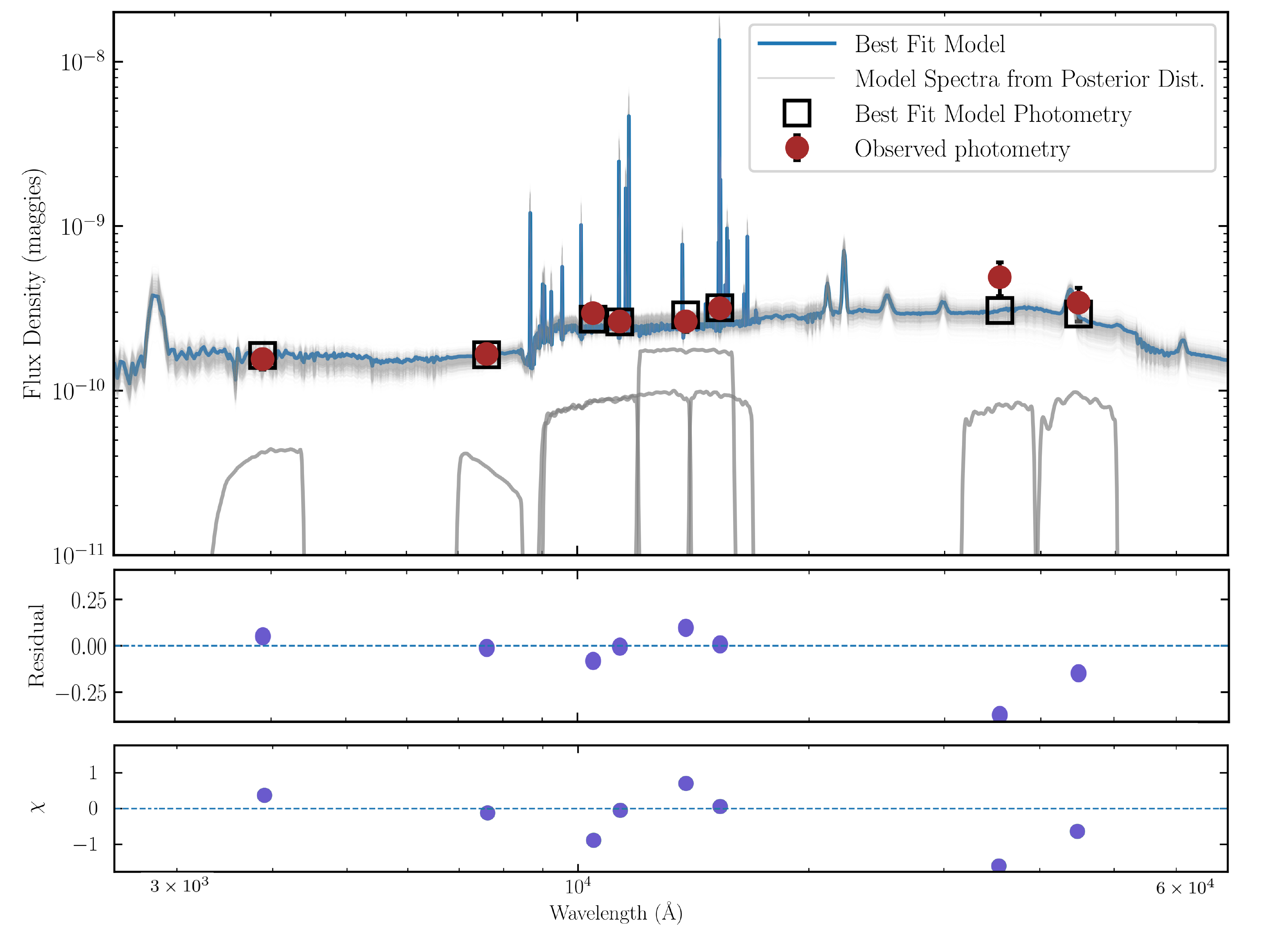}
\caption{Top panel: The best fit spectrum for \seventeen\ is plotted here in blue.  To visualize the spread in model predictions, 1000 other high-likelihood model spectra are plotted in gray.  Photometry from \HST\ and Spitzer data are plotted as red ciricles, with black error bars, while the photometry predicted by the best fit model is plotted as black squares.  For reference, the shapes of the transmission curves for the six \HST\ filters and two Spitzer channels are plotted at the bottom.  Middle: The residual (model minus observed photometry).  Bottom: The $\chi$-value (residual divided by uncertainty) for each point.}
\label{fig:modelSpectra1723}
\end{figure*}

\begin{figure*}[]
\centering
\includegraphics[width=6in]{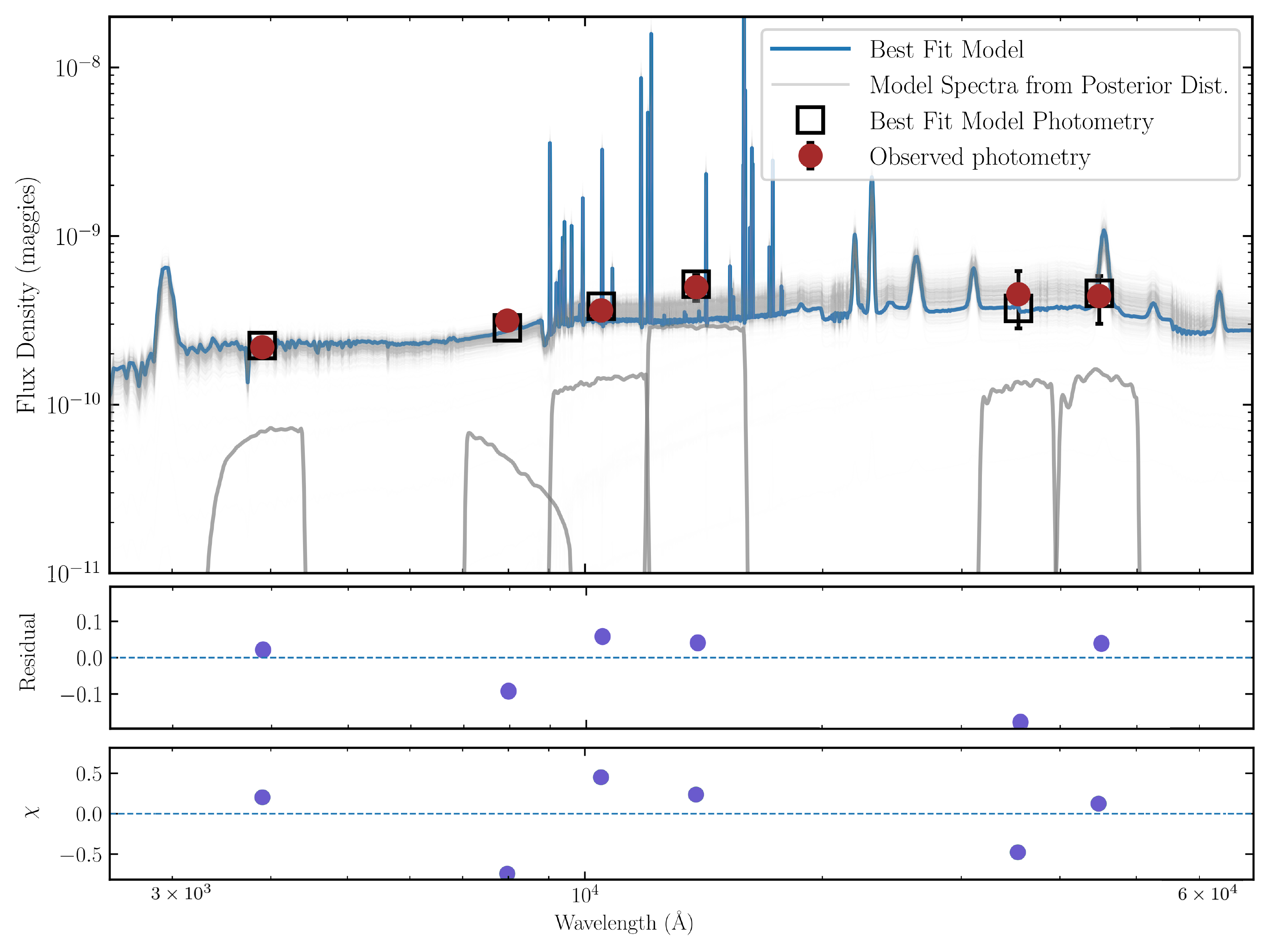}
\caption{The observed photometry, best fit model spectrum, distribution of model spectra, residuals, and $\chi$-values for \twentythree .  Colors and labels follow the same pattern as Fig.~\ref{fig:modelSpectra1723}}
\label{fig:modelSpectra2340}
\end{figure*}

As seen in the corner plots (Fig.~\ref{fig:cornerPlots}), the total mass parameter converges to a tailed Gaussian posterior distribution for these sources, which corresponds to model-generated remnant stellar mass distributions (Figure~\ref{fig:massHists}). The favored model for each of these sources is a recent burst of star formation, corroborated by strong emission features---Ly$\alpha$, H$\alpha$, H$\beta$, H$\gamma$ H$\delta$, OIII[5007]---throughout the best fit model spectra (Fig.~\ref{fig:modelSpectra1723} and  Fig.~\ref{fig:modelSpectra2340} for \seventeen\ and \twentythree\ respectively). 

For \twentythree\, the remnant stellar mass, based on stellar population synthesis (SPS) 
fitting to the 6 photometric data points from HST and Spitzer, is $9.6^{+0.7}_{-0.4} \times 10^{8}$ \Msol . The uncertainties used in the SPS fitting do not include the system-wide 
magnification uncertainty, which when combined with the statistical uncertainties, yields a final mass
of $9.7^{+9.0}_{-4.4} \times 10^{8}$ \Msol .  We should note, however, that the models preferred a dust value slightly outside of the $2\sigma$ constraint inferred from the Balmer decrement.  Relaxing this constraint resulted in the model running to a slightly lower value of the dust2 parameter, but still yielding a stellar mass that agrees within $1\sigma$ ($1.5^{+1.1}_{-0.7} \times 10^{9}$ \Msol ).

\begin{figure*}[]
\centering
\includegraphics[width=7in]{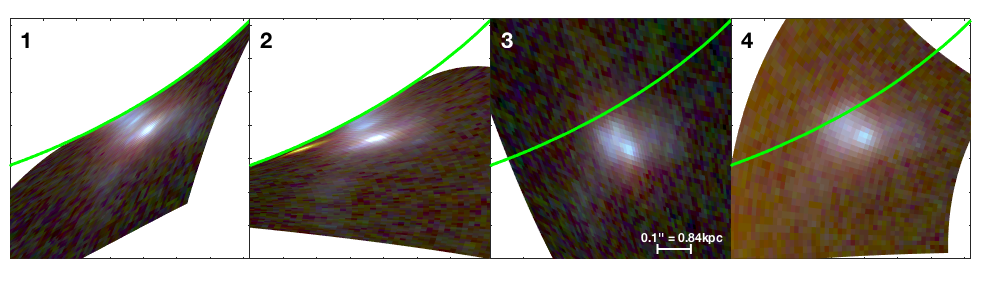}
\caption{Source reconstructions of four of the five images of \seventeen\ labeled as they appear in Fig.~\ref{fig:critCurves}.  Caustics are plotted in green.  We used image 3 to compute the fraction of the source visible in images 1 and 2 in order to correct the stellar mass calculation.}
\label{fig:source1723}
\end{figure*}

For \seventeen , the remnant stellar mass, based on 8 photometric data points from HST and 
Spitzer for the giant arc, is $1.190^{+0.440}_{-0.371} \times 10^{9}$ \Msol .  Because of the excellent constraints on the lens model of \seventeen\, the uncertainties are dominated by statistical uncertainties in the SED fitting rather than by the magnification uncertainties.  The arc, 
however, includes two nearly complete images of the source, so this value must be adjusted by about a factor of two. Figure~\ref{fig:source1723}, shows the location of the caustic in the source reconstruction of images 1 through 4.  Image 3 is a complete image and the exact location of the caustic is better determined for it than for image 4, so we use image 3 to estimate the fraction of the source that is visible in the arc.  To do this, we first drew a custom aperture around the source reconstruction of the stacked UVIS images of image 3.  The UVIS data was chosen because the higher spatial resolution means less flux will be spread across the critical curve by the wings of the PSF.  We then modified that aperture so that it did not cross the caustic (i.e., so that it contained only the portion of the galaxy imaged in the arc).  95.7\% of the flux contained in the original aperture is contained in the modified aperture, so the two images of the source galaxy that make up the giant arc are each about 95.7\% complete.  We therefore adjust the stellar mass estimate by $2 \times 0.957$, or a factor of 1.914.  This results in a stellar mass estimate of 
$5.95^{+2.2}_{-1.86} \times 10^{8}$ \Msol\ for \seventeen .

\begin{table}[h!]
    \label{tab:SM_mags}
    \begin{tabular}{cccc} 
     \toprule
      Source & Band & Magnitude & Uncertainty*\\
      \hline
      \seventeen\ & IRAC Ch1 & 23.28 & 0.25 \\
      & IRAC Ch2 & 23.66 & 0.25 \\
      & F160W & 23.75 & 0.15 \\
      & F140W & 23.95 & 0.15 \\
      & F110W & 23.95 & 0.10 \\
      & F105W & 23.82 & 0.10 \\
      & F775W & 24.44 & 0.10 \\
      & F390W & 24.52 & 0.15 \\

    \hline
    \twentythree\ & IRAC Ch1 & 23.36 & 0.40 \\
      & IRAC Ch2 & 23.39 & 0.34 \\
      & F140W & 23.26 & 0.19 \\
      & F105W & 23.60 & 0.14 \\
      & F814W & 23.76 & 0.14 \\
      & F390W & 24.15 & 0.12 \\

    \end{tabular}
        \caption{Integrated photometry for the arc in \seventeen\ and for the complete image of \twentythree\ determined by combining the three most magnified images.  All values are demagnified. *Uncertainties do not include the magnification uncertainty.}
\end{table}

\clearpage 


\bibliographystyle{astroads}
\bibliography{papers}



\end{document}